\documentclass[prc,notitlepage,twocolumn,showpacs,floatfix,nofootinbib,preprintnumbers,superscriptaddress,amsmath,amssymb,aps,reprint,longbibliography]{revtex4-2}

\usepackage[dvipdfmx]{graphicx}
\usepackage{epsfig}
\usepackage{amsmath,amssymb}
\usepackage{bm}
\usepackage{color}
\usepackage{float}
\usepackage{dcolumn} 
\usepackage{enumerate}
\usepackage{multirow}
\usepackage{threeparttable}
\usepackage{hyperref}
\usepackage{ulem}

\begin{document}

\title{ Global calculation of two-neutrino double-\texorpdfstring{$\beta$}{beta} decay \texorpdfstring{\\}{}
  within the finite amplitude method in nuclear density functional theory}

\author{Nobuo Hinohara}
\email{hinohara@nucl.ph.tsukuba.ac.jp}
\affiliation{
 Center for Computational Sciences, University of Tsukuba, Tsukuba, 305-8577, Japan
}
\affiliation{
 Faculty of Pure and Applied Sciences, University of Tsukuba, Tsukuba, 305-8571, Japan
}
\author{Jonathan Engel}
\email{engelj@physics.unc.edu}
\affiliation{
 Department of Physics and Astronomy,
 University of North Carolina, Chapel Hill,
 North Carolina 27516-3255, USA
}
\date{\today}

\begin{abstract}
Two-neutrino double-beta ($2\nu\beta\beta$) decay has been used to constrain the
neutron-proton part of effective interactions, which in turn is used to compute the
nuclear matrix elements for neutrinoless double-beta decay, the observation of
which would have important consequences for fundamental physics.  We carefully
examine $2 \nu \beta \beta$ matrix elements within the proton-neutron
quasiparticle random-phase approximation with nuclear energy density
functionals.  We work with functionals that are fit globally to single-beta-decay half-lives and charge-exchange giant-resonance energies, but
not to $2\nu \beta \beta$ half-lives themselves, to evaluate the $2\nu \beta
\beta $ nuclear matrix elements for all important nuclei, including those whose
half-lives have not yet been measured.  Such a comprehensive evaluation in large
model spaces without configuration truncation requires an efficient
computational scheme; we employ a double contour integration within the
finite amplitude method.  
The results generally reproduce the nuclear matrix element extracted from
half-lives well, without the use of any of those half-lives in the fitting
procedure.   We present predictions of the matrix elements in a total of 27 nuclei with half-lives that are still unmeasured.
\end{abstract}


\maketitle

\section{Introduction}

Experiments all over the world are attempting to observe neutrinoless
double-beta ($0\nu\beta\beta$) decay, which occurs only if neutrinos are
Majorana particles, at a rate that is related to neutrino masses.  To learn
anything quantitative from an observed decay rate, one must know a nuclear
matrix element that cannot be measured independently and so must be computed
\cite{RevModPhys.80.481,0034-4885-80-4-046301,EJIRI20191,universe6120225}.  Such
computations, which must handle the exchange of a virtual neutrino among
nucleons and mesons, are difficult, and so the matrix elements for isotopes used
in experiments are not known with high precision. A related process,
two-neutrino double-beta ($2\nu\beta\beta$) decay, has been observed, however,
and its rates are often a part of attempts to reduce the uncertainty in
$0\nu\beta\beta$ nuclear matrix elements.

The $2\nu\beta\beta$ nuclear matrix elements have been extracted from measured
half-lives in 11 nuclei at present \cite{universe.6.159}.  To believe the
results of many-body computations of $0\nu\beta\beta$ decay, one would like to
see similar computations that reproduce these $2\nu\beta\beta$ matrix elements.
Because the closure approximation---replacing the energies of states in the
decay's intermediate nucleus with an average---is accurate for
$0\nu\beta\beta$, some approaches rely on it and have a harder time with
$2\nu\beta\beta$ decay, for which the approximation is poor.  Nuclear density
functional theory (DFT) in combination with the proton-neutron
quasiparticle-random phase approximation (pnQRPA) is not one of these
approaches, however; $0\nu\beta\beta$ and $2\nu\beta\beta$ matrix elements can
be computed in similar ways.  In many applications of the pnQRPA, in fact, the
$2\nu\beta\beta$ matrix element is used to constrain the strength of the piece
of the density functional associated with isoscalar proton-neutron pairing,
which suppresses both the $0\nu\beta\beta$ and $2\nu\beta\beta$ matrix elements
\cite{PhysRevLett.57.3148,PhysRevC.37.731}.

The pnQRPA can be used in conjunction with a phenomenological Hamiltonian as
well as in nuclear DFT. The advantage of DFT is its large single-particle model
space and universality; a single energy-density functional (EDF) is taken to
describe all the isotopes in the nuclear chart. The authors of Ref.\
\cite{PhysRevC.87.064302} used a deformed-nucleus pnQRPA with a Skyrme EDF,
computing $2\nu\beta\beta$ and $0\nu\beta\beta$ matrix elements in several
experimentally important isotopes.  They fit the strength of the isoscalar
pairing interaction, on which rates depend sensitively, and it obtains the correct
$2\nu\beta\beta$ matrix elements before computing the $0\nu\beta\beta$ matrix
elements. 

Nuclear EDFs are commonly optimized to reproduce a number of experimental
observables from a wide range of nuclei
\cite{Bogner20132235,PhysRevC.82.024313,PhysRevC.85.024304,
PhysRevC.89.054314,0954-3899-42-3-034024,PhysRevC.79.034310}.  The optimization
is easiest when the observables are ground-state expectation values in even-even
isotopes.  The time-odd part of the EDF
and the proton-neutron pairing strength have no effect on even-even ground states, however, and so
cannot be fixed in the same way.  Instead, they are usually optimized globally,
through the use of single-$\beta$-decay rates and Gamow-Teller and spin-dipole
giant-resonance energies in nuclei all over the table of isotopes
\cite{PhysRevC.93.014304}.

We would like to assess the ability of these globally determined EDFs to
reproduce 2$\nu\beta\beta$ matrix elements so that we can further optimize them if necessary and then confidently apply them to $0\nu\beta\beta$ decay.
Standard pnQRPA calculations, however, require the construction of a QRPA matrix
Hamiltonian, which within large single-particle spaces can consume too much
computational time and memory.  We can turn instead to the finite amplitude
method (FAM) within time-dependent DFT.  The FAM, which is formally equivalent
to the QRPA \cite{nakatsukasa:024318,PhysRevC.84.014314}, computes the linear
response induced by an external field with a complex frequency.  One-body
induced fields and the response of quasiparticle states are calculated by
iteration, without the need to compute the two-body QRPA matrix elements.  The
proton-neutron version of the FAM (pnFAM) was developed and implemented in
Ref.~\cite{PhysRevC.90.024308} in order to calculate $\beta$-decay rates and
Gamow-Teller strength distributions
\cite{PhysRevC.94.055802,PhysRevC.102.034326}.  Because of its efficiency, it
was used in Ref.\ \cite{PhysRevC.93.014304} in an attempt to optimize the
neutron-proton part of a particular nuclear EDF.
   
In this paper we show how to use the pnFAM to efficiently compute
$2\nu\beta\beta$ nuclear matrix elements. Our procedure, a preliminary version
of which was reported on in Ref.~\cite{doi:10.1063/1.5130971}, employs a
complex-plane integration technique \cite{PhysRevC.87.064309,PhysRevC.91.044323}
to perform the summation over intermediate states.  We compare our
$2\nu\beta\beta$ nuclear matrix elements in $^{76}$Ge, $^{130}$Te, $^{136}$Xe,
and $^{150}$Nd to those obtained from matrix diagonalization with the same EDF
in Ref.\ \cite{PhysRevC.87.064302}.  Then we use the EDFs with time-odd terms
fit in Ref.\ \cite{PhysRevC.93.014304} to compute the matrix elements for all 11
nuclei in which the $2\nu\beta\beta$ decay rate has been measured, and for 27 nuclei in which it has not. 

The rest of this paper is organized as follows:  Section~\ref{sec:qrpa} briefly
presents the definition of the $2\nu\beta\beta$ matrix element and describes the
pnQRPA.  Section~\ref{sec:fam} formulates our scheme for computing
$2\nu\beta\beta$ nuclear matrix elements in the pnFAM.  Section~\ref{sec:unc}
compares the pnFAM $2\nu\beta\beta$ matrix elements with those obtained by
matrix diagonalization in the pnQRPA, and Sec.\ \ref{sec:global} assesses the
performance of globally fit functionals and offers predictions for unmeasured
rates.  Section~\ref{sec:conclusion} is a conclusion.

\section{\texorpdfstring{$2\nu\beta\beta$}{2nbb} matrix element and the QRPA\label{sec:qrpa}} 

\subsection{\texorpdfstring{$2\nu\beta\beta$}{2nbb} matrix element}
The nuclear matrix element governing the $2\nu\beta\beta$ decay of the nucleus
$(N,Z)$ to the ground state of the nucleus $(N-2,Z+2)$ contributes to the
half-life $T^{2\nu}_{1/2}$ as follows:
\begin{equation}
\label{eq:T2nu}
[T^{2\nu}_{1/2}]^{-1} = G_{2\nu}(Q_{\beta\beta},Z) |M^{2\nu}|^2 \,,
\end{equation}
where $G_{2\nu}$ is a phase space factor, and the 2$\nu\beta\beta$ matrix
element is a sum of Fermi and Gamow-Teller parts \cite{RevModPhys.80.481},
\begin{align}
  M^{2\nu} &= M^{2\nu}_{\rm GT} - \frac{g_V^2}{g_A^2} M^{2\nu}_{\rm F}, \\
  M^{2\nu}_{\rm F}&= \sum_n \frac{ \displaystyle\langle 0^+_f|\sum_a \tau^-_a |n\rangle
    \langle n|\sum_b \tau^-_b |0^+_i\rangle}
  {\displaystyle E_n - \frac{M_i+M_f}{2}}, \\
  M^{2\nu}_{\rm GT}&= \sum_n \frac{ \displaystyle\langle 0^+_f|\sum_a \bm{\sigma}_a \tau^-_a |n\rangle
  \cdot  \langle n|\sum_b \bm{\sigma}_b \tau^-_b |0^+_i\rangle}
  {\displaystyle E_n - \frac{M_i+M_f}{2}}.
  \label{eq:2nGT}
\end{align}
Here $\tau^-_a$ is the isospin-lowering operator for nucleon $a$,
$\bm{\sigma}_a$ is the corresponding spin operator, $M_i$ and $M_f$ are the
ground-state energies of the initial and final states of the decay, and $|n\rangle$, with energy $E_n$ is one of a
complete set of intermediate states in the nucleus $(N-1,Z+1)$.  The Fermi part
of the $2\nu\beta\beta$ matrix element is very small because isospin is nearly
conserved \cite{PhysRevC.87.045501}, and we neglect it here. 

\subsection{The pnQRPA}

The proton-neutron QRPA evaluates the transition matrix elements between the
initial or final state and the intermediate states that appear in the numerator
of Eq.~(\ref{eq:2nGT}), taking into account the effect of the proton-neutron
residual interaction beyond the mean-field approximation.  In the pnQRPA, both
the initial and final states $|0_{i/f, {\rm QRPA}}^+\rangle$ are based on
Hartree-Fock-Bogoliubov (HFB) quasiparticle vacua, which incorporate
axially-symmetric deformation in our work.  The intermediate states are related
to the initial or final state by a QRPA phonon operator
\begin{equation}
\begin{aligned}
|\lambda,K\rangle &= 
\hat{\cal Q}^{\lambda\dag}_{K}
|0^+_{{\rm QRPA}}\rangle \\
\hat{\cal Q}^{\lambda\dag}_{K}
&= \!\!\!\!\!\! \sum_{\substack{pn \\ j_{z,p}+j_{z,n}=K}}
\!\!\!\! \!\! X_{pn,K}^{\lambda} \hat{a}^{\dag}_{p} \hat{a}^{\dag}_{n}
- Y_{pn,K}^{\lambda} 
\hat{a}_{\bar{n}} 
\hat{a}_{\bar{p}} \,,
\end{aligned}
\end{equation}
where $\hat{a}_{\tau=n,p}$ is a neutron or proton quasiparticle operator,
defined so that $\hat{a}_{\tau}|0_{\rm HFB}^+\rangle = 0$.  Here, the indices
$p$ and $n$ label proton and neutron quasiparticles.  $j_{z,\tau}$ and $K$ are
the projections along the symmetry axis of the quasiparticle and phonon angular
momentum, and the index $\bar{\tau}$ labels the time-reversal partner of the
state $\tau$ ($j_{z,\bar{\tau}} = -j_{z,\tau}$).  From now on, for the sake of
simplicity we omit the restriction $j_{z,p}+j_{z,n}=K$ when summing over the
proton and neutron quasiparticle states.

The QRPA amplitudes $X_{pn,K}^\lambda$ and $Y_{pn,K}^\lambda$ are solutions of
the QRPA equations,
\begin{align}
  \sum_{p'n'}    
  \begin{pmatrix} A_{pn,p'n'} & B_{pn,p'n'} \\ B^{\ast}_{pn,p'n'} & A^{\ast}_{pn,p'n'} \end{pmatrix}
  \begin{pmatrix} X^{\lambda}_{p'n',K} \\ Y^{\lambda}_{p'n',K} \end{pmatrix}
    =
    \Omega^{\lambda}_{K}
    \begin{pmatrix} X^{\lambda}_{pn,K} \\ -Y^{\lambda}_{pn,K} \end{pmatrix},
\end{align}
where $\Omega^{\lambda}_{K}$ is an excitation energy, measured from the QRPA
ground state of the initial/final state.  The $A$ and $B$ matrices contain
residual interactions, computed from the second functional derivative of the
EDF.  The $2\nu\beta\beta$ matrix element can be calculated by combining the
pnQRPA transition matrix elements from the initial and final states of the decay
to the intermediate states.  Because the procedure introduces two sets of the
intermediate states, an additional approximation for matching them is necessary.
We thus approximate the Gamow-Teller matrix element in Eq.~(\ref{eq:2nGT}) by 
\begin{widetext}
\begin{align}
  M^{2\nu}_{\rm GT} &= \sum_{K=-1}^1 (-1)^K \sum_{\substack{\lambda_i>0 \\ \lambda_f>0}} \frac{\displaystyle
  \langle 0_{f,{\rm QRPA}}^+| \hat{F}^{{\rm GT}-}_{-K} |\lambda_f,K\rangle
    \langle \lambda_f,K|\lambda_i,K\rangle \langle \lambda_i,K|\hat{F}^{{\rm  GT}-}_{K} |0_{i,{\rm QRPA}}^+\rangle}
      {\displaystyle \frac{\Omega^{\lambda_i}_{K} + \Omega^{\lambda_f}_{K}}{2}}.
      \label{eq:2nGTQRPA}
\end{align}
\end{widetext}
In the summation, the expression $\lambda>0$ denotes the states with
$\Omega_K^\lambda>0$. 

The Gamow-Teller operator in the quasiparticle basis is 
\begin{align}
    \hat{F}^{{\rm GT}\pm}_{K} &= \sum_a (\sigma_K)_a \tau^\pm_a \nonumber\\
    &= \sum_{pn} \left[F_{20,K}^{{\rm GT}\pm}(pn) \hat{a}^\dag_{p} \hat{a}^\dag_{n}
     + F_{02,K}^{{\rm GT}\pm}(pn) \hat{a}_{\bar{n}} \hat{a}_{\bar{p}}\right] \nonumber \\
    &\quad  + ( \hat{a}^\dag \hat{a}{\rm -terms}),
\end{align}
and its transition amplitudes in Eq. (\ref{eq:2nGTQRPA}) are given by
\begin{equation}
\begin{aligned}
\langle \lambda_i, K| 
\hat{F}^{{\rm GT-}}_{K}|0^+_{i,{\rm QRPA}}\rangle=
& \\ 
\sum_{pn} \left[F^{{\rm GT-}}_{20,K}(pn) X^{\lambda_i\ast}_{pn,K} 
\right.&\left.+ F^{{\rm GT-}}_{02,K}(pn) Y^{\lambda_i\ast}_{pn,K} \right] \\
\langle  0^+_{f,{\rm QRPA}}| 
\hat{F}^{{\rm GT-}}_{-K}|\lambda_f,K\rangle=
& \\ 
\sum_{pn}\left[ F^{{\rm GT-}}_{02,K}(pn) X^{\lambda_f}_{pn,K}
\right.&\left.+ F^{{\rm GT-}}_{20,-K}(pn) Y^{\lambda_f}_{pn,-K}\right] \,. 
\end{aligned}
\end{equation}
To compute the overlap of the two intermediate states $\langle
\lambda_f,K|\lambda_i,K\rangle$ we adapt expressions based on the QRPA
\cite{Simkovic2004321} and the quasiparticle Tamm-Dancoff approximation (QTDA)
\cite{PhysRevC.87.064302}.  The result is
\begin{align}
\langle \lambda_f,K|\lambda_i,K\rangle
&= \sum_{pnp'n'}
\left( X_{p'n',K}^{\lambda_f\ast} X_{pn,K}^{\lambda_i}
- \alpha Y_{p'n',K}^{\lambda_f\ast} Y_{pn,K}^{\lambda_i}
\right) \nonumber \\
& \quad \times{\cal O}_{pp'}(\alpha) {\cal O}_{nn'}(\alpha)
\nonumber\\
&= \sum_{pn}
\left( \bar{X}_{pn,K}^{\lambda_f\ast} \bar{X}_{pn,K}^{\lambda_i}
- \alpha \bar{Y}_{pn,K}^{\lambda_f\ast} \bar{Y}_{pn,K}^{\lambda_i}
\right), \label{eq:overlap1}
\end{align}
where $\alpha$ is a parameter that is 0 for the QTDA overlap and 1 for the QRPA
overlap, and the ${\cal O}_{\tau\tau'}(\alpha)$ are elements of the matrix that
connect the quasiparticles associated with the initial and final states of the
decay.  Explicit expressions for these elements, together with the derivation of
Eq.~(\ref{eq:overlap1}), are in Appendix~\ref{sec:overlap}.  $\bar{X}$ and
$\bar{Y}$ are defined by
\begin{subequations}\label{eq:XbarYbar}
\begin{align}
\bar{X}_{pn,K}^{\lambda_i} &=\sum_{p'} {\cal O}^{T}_{pp'}(\alpha) 
X_{p'n,K}^{\lambda_i},\\
\bar{Y}_{pn,K}^{\lambda_i} &= \sum_{p'} {\cal O}^{T}_{pp'}(\alpha) Y_{p'n,K}^{\lambda_i},\\
\bar{X}_{pn,K}^{\lambda_f} &= \sum_{n'} X_{pn',K}^{\lambda_f}{\cal O}^{T}_{n'n}(\alpha), \\
\bar{Y}_{pn,K}^{\lambda_f} &= \sum_{n'} Y_{pn',K}^{\lambda_f}{\cal O}^{T}_{n'n}(\alpha).
\end{align}
\end{subequations}

\section{The FAM \label{sec:fam}}
\subsection{pnFAM}

The FAM is formally equivalent to the QRPA and enables us to compute DFT
response functions efficiently.  A detailed formulation of the like-particle FAM
and the pnFAM in the presence of the pairing correlations appear, respectively,
in Refs.~\cite{PhysRevC.84.014314} and \cite{PhysRevC.90.024308}.

In the pnFAM, one applies a time-dependent external field of the form
\begin{align}
\hat{F}_{K}^T(t) = \eta( \hat{F}_{K}^T e^{i\omega t} + \hat{F}^{T\dag}_{K}
e^{-i\omega t}) \,, 
\end{align}
with $\hat{F}^T_K$ a one-body proton-neutron excitation operator and $\omega$ a
complex frequency.  The excitation operator induces oscillations of
quasiparticle annihilation operators (e.g., for neutrons) of the form 
\begin{align}
\delta \hat{a}_n(t) &= \eta \sum_p \hat{a}^\dag_p \left[
X_{pn}(\omega,\hat{F}^T_K) e^{-i\omega t} + Y^\ast_{pn}(\omega,\hat{F}^T_K)
e^{i\omega t} \right].
\end{align}
Solving the time-dependent DFT equations results in the FAM amplitudes
$X_{pn}(\omega,\hat{F}^T_K)$ and $Y_{pn}(\omega,\hat{F}^T_K)$, which are related
to the QRPA amplitudes $X^\lambda_{pn,K}$ and $Y^\lambda_{pn,K}$ through
\cite{PhysRevC.87.064309}
\begin{align}
  X_{pn}(\omega,\hat{F}_{K}^T)&= - \sum_{\lambda>0} \left\{
  \frac{X_{pn,K}^{\lambda} \langle \lambda,K|\hat{F}_{K}^T|0^+\rangle}
  {\Omega^\lambda_K- \omega} \right. \nonumber \\
    &\quad + \left.  \frac{ Y_{pn,K}^{\lambda\ast}\langle
    0^+|\hat{F}_{K}^T|\lambda,-K\rangle} {\Omega^\lambda_K+\omega}
    \right\}, \label{eq:FAMX}\\
    Y_{pn}(\omega,\hat{F}_{K}^T)&= - \sum_{\lambda>0} \left\{
    \frac{Y_{pn,K}^{\lambda} \langle \lambda,K|\hat{F}_{K}^T|0^+\rangle}
    {\Omega^\lambda_K- \omega} \right.\nonumber \\
    &\quad + \left.  \frac{ X_{pn,K}^{\lambda\ast}\langle
    0^+|\hat{F}_{K}^T|\lambda,-K\rangle} {\Omega^\lambda_K+\omega}
    \right\}. \label{eq:FAMY}
\end{align}

\subsection{\texorpdfstring{$2\nu\beta\beta$}{2nbb} matrix elements in the pnFAM}

To calculate the QRPA $2\nu\beta\beta$ nuclear matrix element in Eq.\
\eqref{eq:2nGTQRPA}, we separately solve the pnFAM computations in the initial
and final nuclei, distinguishing quantities from the two nuclei with the
superscripts $(i)$ and $(f)$.  We then compute a quantity that is a combination
of the two sets of pnFAM amplitudes
\begin{align}
  {\cal T}(\alpha;\omega_i, \hat{F}_{K_i}^{T_i}; \omega_f, \hat{F}_{K_f}^{T_f})
  & \equiv \sum_{pn} \left[
        \bar{Y}^{(f)}_{pn}(\omega_f,\hat{F}_{K_f}^{T_f})
        \bar{X}^{(i)}_{pn}(\omega_i,\hat{F}_{K_i}^{T_i})
        \right. \nonumber \\
& \quad\left. -\alpha \bar{X}^{(f)}_{pn}(\omega_f,\hat{F}_{K_f}^{T_f})
        \bar{Y}^{(i)}_{pn}(\omega_i,\hat{F}_{K_i}^{T_i}) \right],
        \label{eq:Tfunc}
\end{align}
\begin{figure}[t]
\includegraphics[width=\columnwidth]{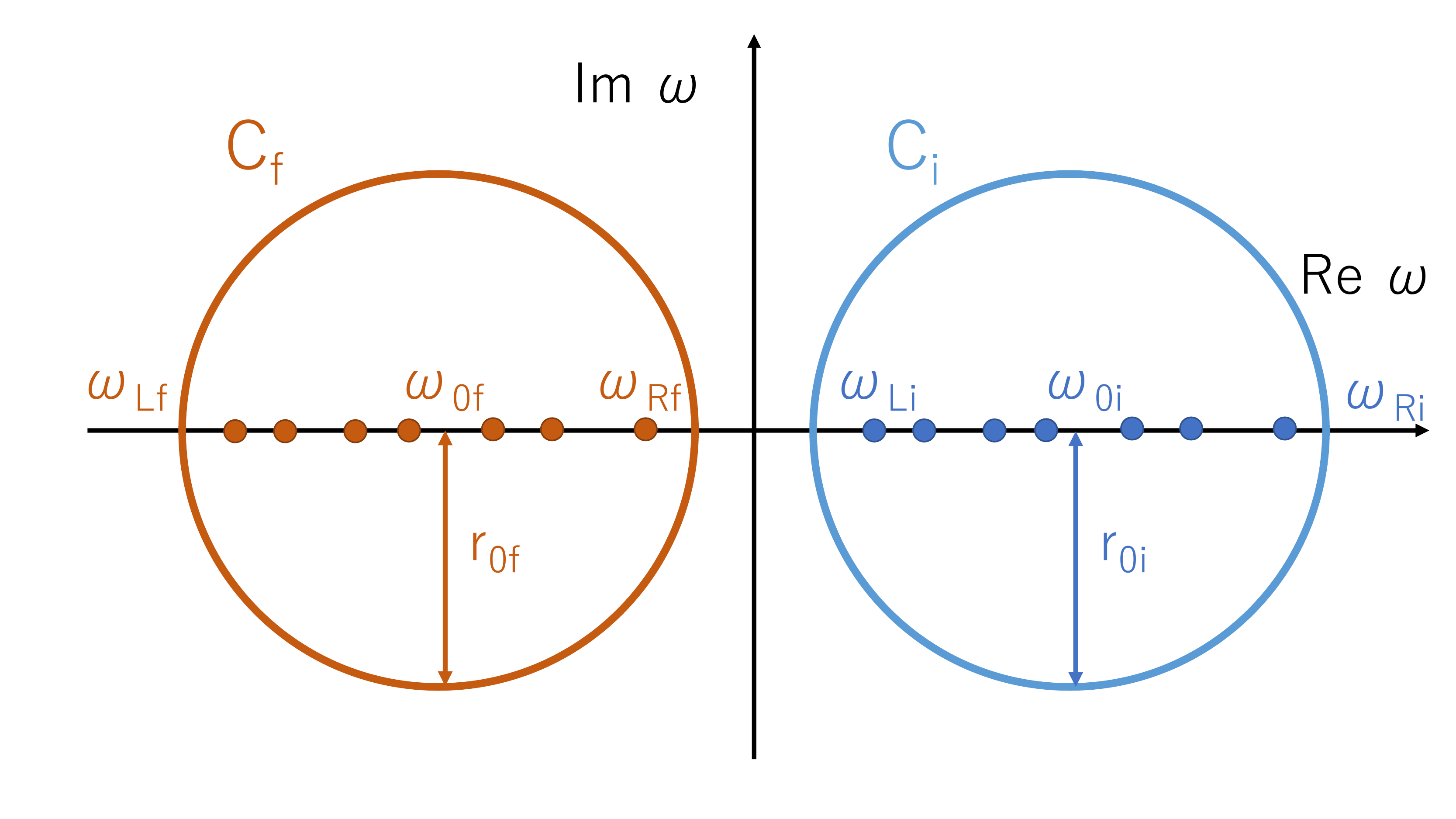}
\caption{Contours $C_i$ and $C_f$. \label{fig:contour}}
\end{figure}
where $\bar{X}^{(i/f)}$ and $\bar{Y}^{(i/f)}$ are the amplitudes in
Eqs.~(\ref{eq:FAMX}) and (\ref{eq:FAMY}), labeled in the same way as the QRPA
amplitudes in Eq.~(\ref{eq:XbarYbar}).  By substituting Eqs.~(\ref{eq:FAMX}) and
(\ref{eq:FAMY}) into Eq.~(\ref{eq:Tfunc}), we obtain an expression for
$\mathcal{T}$ in terms of the QRPA amplitudes: 
\begin{widetext}
\begin{align}
  {\cal T}(\alpha;\omega_i,\hat{F}_{K_i}^{T_i}; \omega_f, \hat{F}_{K_f}^{T_f})
  &=
  \sum_{pn}
  \sum_{\substack{\lambda_i>0 \\ \lambda_f>0}} \left[
    \frac{ (
    \bar{X}^{\lambda_f\ast}_{pn,K_f} \bar{X}^{\lambda_i}_{pn,K_i} - \alpha     \bar{Y}^{\lambda_f\ast}_{pn,K_f} \bar{Y}^{\lambda_i}_{pn,K_i})
    \langle 0_f^+|\hat{F}_{K_f}^{T_f}|\lambda_f,-K_f\rangle \langle \lambda_i,K_i|\hat{F}_{K_i}^{T_i}|0_i^+\rangle}
       { (\Omega^{\lambda_f}_{K_f} + \omega_f)(\Omega^{\lambda_i}_{K_i} - \omega_i)} \right. \nonumber \\
    &\quad +
   \frac{ (
   \bar{Y}^{\lambda_f\ast}_{pn,K_f}
   \bar{X}^{\lambda_i}_{pn,K_i} - \alpha 
   \bar{X}^{\lambda_f\ast}_{pn,K_f} 
   \bar{Y}^{\lambda_i}_{pn,K_i})
  \langle \lambda_f,K_f|\hat{F}_{K_f}^{T_f}|0_f^+\rangle \langle \lambda_i,K_i|\hat{F}_{K_i}^{T_i}|0_i^+\rangle}
              { (\Omega^{\lambda_f}_{K_f} - \omega_f)(\Omega^{\lambda_i}_{K_i} - \omega_i)} \nonumber \\
    &\quad +
    \frac{ (
     \bar{X}^{\lambda_f\ast}_{pn,K_f} 
     \bar{Y}^{\lambda_i}_{pn,K_i} - \alpha \bar{Y}^{\lambda_f\ast}_{pn,K_f} 
     \bar{X}^{\lambda_i}_{pn,K_i})
  \langle 0_f^+|\hat{F}_{K_f}^{T_f}|\lambda_f,-K_f\rangle \langle 0_i^+|\hat{F}_{K_i}^{T_i}|\lambda_i,-K_i\rangle}
   { (\Omega^{\lambda_f}_{K_f} + \omega_f)(\Omega^{\lambda_i}_{K_i} + \omega_i)} \nonumber \\
    &\quad + \left.
       \frac{ (
       \bar{Y}^{\lambda_f\ast}_{pn,K_f} 
       \bar{Y}^{\lambda_i}_{pn,K_i} - \alpha \bar{X}^{\lambda_f\ast}_{pn,K_f} 
       \bar{X}^{\lambda_i}_{pn,K_i})
  \langle \lambda_f,K_f|\hat{F}_{K_f}^{T_f}|0_f^+\rangle \langle 0_i^+|\hat{F}_{K_i}^{T_i}|\lambda_i,-K_i\rangle}
  { (\Omega^{\lambda_f}_{K_f} - \omega_f)(\Omega^{\lambda_i}_{K_i} + \omega_i)} \right]. \label{eq:T}
\end{align}
\end{widetext}
${\cal T}$ has first-order poles at $\omega_i = \pm\Omega^{\lambda_i}_{K_i}$ and
$\omega_f=\pm\Omega^{\lambda_f}_{K_f}$.  We choose a counterclockwise contour
$C_i$ ($C_f$) for $\omega_i$ ($\omega_f$) that includes positive-energy
(negative-energy) poles and excludes all the negative-energy (positive-energy)
poles, as in Fig.~\ref{fig:contour}, to extract the residues from the first term
on the right side of Eq.~(\ref{eq:T}).

Integrating $\mathcal{T}$ together with an arbitrary regular complex function of
$\omega_i$ and $\omega_f$ over those frequencies, we obtain an expression that
can be used for two-body matrix elements:
\begin{align}
  &{\cal M}\bm{\bigl(}\alpha; \hat{F}_{K_i}^{T_i}, \hat{F}_{K_f}^{T_f}; f(\omega_i,\omega_f)\bm{\bigr)} \nonumber \\
  &=-\frac{1}{4\pi^2}
  \oint_{C_i} d\omega_i \oint_{C_f} d\omega_f {\cal T}(\alpha;\omega_i,\hat{F}_{K_i}^{T_i};\omega_f,\hat{F}_{K_f}^{T_f})
  f(\omega_i,\omega_f) \nonumber \\
  &= \sum_{\substack{\lambda_i>0\\ \lambda_f>0}}
  \sum_{pn}
  \left(\bar{X}^{\lambda_f\ast}_{pn,K_f} \bar{X}^{\lambda_i}_{pn,K_i} - \alpha \bar{Y}^{\lambda_f\ast}_{pn,K_f} \bar{Y}^{\lambda_i}_{pn,K_i} \right) \nonumber \\
  &\quad \times
  f(\Omega^{\lambda_i}_{K_i},-\Omega^{\lambda_f}_{-K_f}) 
  \langle 0_f^+|\hat{F}_{K_f}^{T_f}|\lambda_f,-K_f\rangle \langle \lambda_i,K_i|\hat{F}_{K_i}^{T_i}|0_i^+\rangle.
  \label{eq:M}
\end{align}
The Fermi and Gamow-Teller $2\nu\beta\beta$ decay nuclear matrix elements are
then given by 
\begin{align}
M^{2\nu}_{{\rm F}} &= {\cal M}\left(\alpha; \hat{F}^{{\rm F-}},\hat{F}^{{\rm F-}}; f=\frac{2}{\omega_i - \omega_f}\right), \\
  M^{2\nu}_{{\rm GT}} &= \sum_{K=-1}^1 (-1)^K {\cal M}\left(\alpha; \hat{F}_K^{{\rm GT-}},\hat{F}_{-K}^{{\rm GT-}}; f=\frac{2}{\omega_i - \omega_f}\right),
\end{align}
under the assumptions that $X^{\lambda_f}_{-K}=X^{\lambda_f}_K$,
$Y^{\lambda_f}_{-K}=Y^{\lambda_f}_{K}$, and $\Omega^{\lambda_f}_{-K_f} =
\Omega^{\lambda_f}_{K_f}$.  Even when starting from the final state we use the
external operator $\sigma_{-K}\tau^-$ that changes neutrons into protons to
properly include the backward amplitudes in Eqs.~(\ref{eq:FAMX}) and
(\ref{eq:FAMY}). 

By setting $f=1$, $\alpha=1$, and taking the same HFB vacuum for the initial and
final states in Eq.~(\ref{eq:M}), we can use that equation to compute the
unweighted summed strengths: 
\begin{align}
  {\cal M}^{i=f}(1;\hat{F}^{\rm F\mp}, \hat{F}^{\rm F\pm}; 1)
 &= \sum_{\lambda>0} |\langle \lambda,0|\hat{F}^{\rm F\mp}|0^+\rangle|^2, \\
  {\cal M}^{i=f}(1; \hat{F}^{\rm GT\mp}_{K}, \hat{F}^{\rm GT\pm}_{-K};1)
 &= (-1)^K \nonumber \\ 
 &\quad \times \sum_{\lambda>0} |\langle \lambda,K|\hat{F}^{{\rm GT\mp}}_{K}|0^+\rangle|^2.
\end{align}
Sum rules can be used to check the routines that compute matrix elements. 

\section{Results with SkM* and Comparison with Prior Work \label{sec:unc}}

Our calculation of $2\nu\beta\beta$ nuclear matrix elements uses an extension of
the pnFAM code developed in Ref.~\cite{PhysRevC.90.024308}, which is in turn
based on the nuclear DFT solver {\sc hfbtho}
\cite{PEREZ2017363,Stoitsov20131592,Stoitsov200543}.  That last code uses the
harmonic oscillator basis in a cylindrical coordinate system and allows axial
deformation.  In this section we provide details of our calculations with the
SkM* functional and compare our $2\nu\beta\beta$ matrix elements for $^{76}$Ge,
$^{130}$Te, $^{136}$Xe, and $^{150}$Nd with those obtained in
Ref.~\cite{PhysRevC.87.064302} by diagonalizing the pnQRPA matrix.
\subsection{Parameter values}

\begin{table}[t]
 \caption{Experimental values of $\tilde{\Delta}_n^{(3)}$ and
 $\tilde{\Delta}_p^{(3)}$ (in MeV) and the volume-pairing strengths $V_n$ and
 $V_p$ fit to those values (in MeV fm$^3$).  The averages of the strengths in
 the initial and final nuclei used in the pairing EDF.  Experimental binding
 energies are taken from Ref.~\cite{ame2020}.  Data in parentheses are not used
 to fit the pairing strengths.
  \label{table:gap}}
 \begin{ruledtabular}
   \begin{tabular}{ccccc} 
    & $\tilde{\Delta}_n^{(3)}$ & $\tilde{\Delta}_p^{(3)}$ & $V_n$ & $V_p$
      \\ \hline
     $^{76}$Ge & 1.393 & 1.114 & $-$182.70 & $-$194.49 \\
     $^{76}$Se & 1.551 & 1.392 & $-$185.40 & $-$202.22 \\
     average && & $-$184.05 & $-$198.36 \\ \hline
     $^{130}$Te& 1.114 & (0.801)& $-$166.21 & N/A \\
     $^{130}$Xe& 1.170 & 1.016 &$-$173.80 & $-$194.00 \\
     average & & & $-$170.01 & $-$194.00 \\ \hline
     $^{136}$Xe& (0.841) & 0.751 & N/A & $-$148.66 \\
     $^{136}$Ba& 0.960 & 1.005 &$-$184.16 & $-$172.54 \\
     average & & & $-$184.16 & $-$160.60 \\ \hline
     $^{150}$Nd& 1.070 & 0.918 & $-$181.64 & $-$202.31 \\
     $^{150}$Sm& 1.194 & 1.196 & $-$184.84 & $-$195.24 \\
     average & & & $-$183.24 & $-$198.78
     \end{tabular}
  \end{ruledtabular}
\end{table}

To integrate in cylindrical coordinates, we use Gauss-Hermite quadrature with
$N_{\rm GH}=40$ points for the $z$ direction and Gauss-Laguerre quadrature with
$N_{\rm GL}=40$ points for the $r$ direction.  To compute the direct Coulomb
mean field, we use the prescription described in Ref.~\cite{Stoitsov20131592}
with length parameter $L=50$ fm and $N_{\rm Leg}=80$ Gauss-Legendre points. 

We include $N_{\rm sh}=20$ harmonic-oscillator major shells to describe the HFB
wave functions. This corresponds to 1771 single-particle states for neutrons
and protons (with axial and time-reversal symmetry taken into account), and, in
the pnFAM, to 257 686 $K=0$ two-quasiparticle states and 256 025 $K=\pm 1$
two-quasiparticle states.  We include all such states, with no additional
model-space truncation, in the pnFAM calculations.  The dimension of the pnQRPA
matrix corresponding to our pnFAM calculations is about 500 000 for each $K$
quantum number.

We employ the same Skyrme SkM* functional \cite{Bartel198279} and volume-type
pairing with 60-MeV energy cutoff (with $\hbar^2/2m=20.73$ MeV fm$^2$ and the
one-body center-of-mass correction included in the kinetic term) as that in
Ref.\ \cite{PhysRevC.87.064302}.  The HFB solver cited in that paper, however,
works in a cylindrical box with $r_{\rm max}=z_{\rm max}=20$ fm and a coordinate
spacing of 0.7 fm, and is thus different from ours.

\begin{table}[b]
  \caption{Properties of HFB ground states with the SkM* + volume pairing (with
  average pairing strengths) EDF. The table shows pairing gaps (in MeV),
  quadrupole deformation, and total HFB energies (in MeV) and compares the
  quadrupole deformation to the value in Ref.~\cite{PhysRevC.87.064302}. 
\label{table:dft}}
  \begin{ruledtabular}
    \begin{tabular}{cccccc}
      & \textrm{$\Delta_n$} & \textrm{$\Delta_p$} & \textrm{$\beta$} &
      \textrm{$E_{\rm HFB}$} & \textrm{$\beta$ (Ref \cite{PhysRevC.87.064302})}
      \\
      \colrule $^{76}$Ge & 1.609 & 1.473 & $-$0.021 & $-$661.804 & $-$0.025 \\
     & 1.435 & 1.205 & 0.185 & $-$662.274 & 0.184 \\
     & 1.612 & 1.475 & 0. & $-$661.802 & \\ \hline
      $^{76}$Se & 1.589 & 1.648 & 0. & $-$659.315 & $-$0.018 \\
     & 1.508 & 1.257 & $-$0.194 & $-$659.594 & \\ \hline
     $^{130}$Te 
     & 1.178 & 1.028 & 0. & $-$1096.839 & 0.01 \\ \hline
     $^{130}$Xe & 1.078 & 1.009 & 0.141 & $-$1093.423 & 0.13 \\
     & 1.107 & 1.113 & $-$0.124 & $-$1093.152 & \\
     & 1.359 & 1.351 & 0. & $-$1092.393 & \\ \hline
      $^{136}$Xe 
     & 0. & 0.878 & 0. & $-$1143.253 & 0.004 \\ \hline
      $^{136}$Ba & 1.025 & 0.931 & $-$0.047 & $-$1139.268 & $-$0.021\\
     & 0.928 & 0.735 & 0.094 & $-$1139.538 & \\
     & 1.057 & 0.985 & 0. & $-$1139.231 & \\ \hline
      $^{150}$Nd & 1.129 & 0.764 & 0.292 & $-$1235.794 & 0.27 \\
      & 1.375 & 1.358 & $-$0.177 & $-$1232.563 & \\
      & 1.422 & 1.688 & 0. & $-$1231.080 & \\ \hline
      $^{150}$Sm & 1.131 & 1.307 & 0.223 & $-$1234.675 & 0.22 \\ 
     & 1.294 & 1.707 & 0. & $-$1232.436 & \\
     & 1.305 & 1.534 & $-$0.137 & $-$1233.068 & 
    \end{tabular}
  \end{ruledtabular}
  \end{table}

We adjust the volume pairing strengths to reproduce experimental odd-even
staggering (OES) with the density-averaged pairing gap.  To reduce fluctuations
\cite{PhysRevC.82.024313}, we take as the experimental data an average of the
results of the three-point formula evaluated at the two even-odd or odd-even
systems:
\begin{equation}
\begin{aligned}
\tilde{\Delta}^{(3)}_n(N,Z) &= \frac{ \Delta^{(3)}_n(N-1,Z) +
\Delta^{(3)}_n(N+1,Z)}{2} \\
\tilde{\Delta}^{(3)}_p(N,Z) &= \frac{\Delta^{(3)}_p(N,Z-1) +
\Delta^{(3)}_p(N,Z+1)}{2} \,, 
\end{aligned}
\end{equation}
where $\Delta^{(3)}_{n/p}$ is the result of the three-point formula \cite{Bender2000}.
Table \ref{table:gap} lists the experimental values for this quantity and the
neutron and proton volume pairing strengths that best reproduce them.   In order
to use the same EDF for both nuclei in the decay, we take the average of the
pairing strengths fit in the initial and final nuclei.  We note that the
experimental $\Delta^{(3)}_{n/p}$ values do not provide useful information if
the series of isotopes used to calculate them includes closed-shell nuclei.
$\Delta^{(3)}_{p}$ in $^{130}$Te ($Z=50$ included) and $\Delta^{(3)}_{n}$ in
$^{136}$Xe and $^{136}$Ba ($N=82$ included) are such cases if the average of the
results of two odd-even mass formulas is used.  We avoid using the pairing gap
$\tilde{\Delta}_p^{(3)}$ of $^{130}$Te to fit the proton pairing strength,
fitting the pairing strength instead to the proton gap in $^{130}$Xe.  We do
adopt the neutron $\tilde{\Delta}^{(3)}$ of $^{136}$Ba, however, to determine
the neutron pairing strength because the strengths fit to $\Delta_n^{(3)}$ and
$\tilde{\Delta}_n^{(3)}$ are quite similar in that nucleus.  The globally fit
EDFs described in Sec.~\ref{sec:global} are free from these problems. 

Table \ref{table:dft} shows the results of the DFT calculations for the initial
and final nuclei.  The quadrupole deformations of the HFB states are quite close
to those in Ref.~\cite{PhysRevC.87.064302}.  We choose the HFB solution in the
top line for each nucleus in which several HFB solutions coexist.

\begin{table}[t]
  \caption{$Q$ values for each double-beta decay in units of MeV.  Experimental
  $Q$ values are obtained from atomic masses \cite{ame2020}.
\label{table:qvalue}}
  \begin{ruledtabular}
    \begin{tabular}{ccccc}
      & This paper & SkM* (Ref.~\cite{PhysRevC.87.064302}) & Exp. \\ \colrule
      $^{76}$Ge $\to$ $^{76}$Se &  4.05 & 4.84 & 2.04 \\
      $^{130}$Te $\to$ $^{130}$Xe & 4.98 & 4.22 & 2.53 \\
      $^{136}$Xe $\to$ $^{136}$Ba & 5.55 & 5.60 & 2.46 \\
      $^{150}$Nd $\to$ $^{150}$Sm & 2.68 & 2.35 & 3.37 \\
    \end{tabular}
  \end{ruledtabular}
  \end{table}

Table~\ref{table:qvalue} shows $\beta\beta$ $Q$ values.  Our calculation does
not perfectly reproduce the values in Ref.~\cite{PhysRevC.87.064302}, which were obtained
from the same SkM* EDF but a different HFB code.  We suspect that the
differences are due to the different representations of the oscillator basis
states and treatments of pairing.

\begin{table}[b]
\caption{Neutron and proton parts of the HFB overlap $\langle 0^+_{f,{\rm
HFB}}|0^+_{i,{\rm HFB}}\rangle$ between the initial and the final states
compared with values from previous QRPA calculations.
\label{table:overlap}}
\begin{ruledtabular} 
  \begin{tabular}{cccccccccc}
     & neutron & proton & total & Ref.~\cite{PhysRevC.87.064302} &
     Ref.~\cite{PhysRevC.83.034320} & Ref.~\cite{PhysRevC.97.045503} \\ \colrule
    $^{76}$Ge & 0.907 & 0.886 & 0.803 & & 0.81 & 0.72, 0.73\\
    $^{130}$Te & 0.329 & 0.403 & 0.133 & & & 0.73, 0.73\\
    $^{136}$Xe & 0.480 & 0.787 & 0.378 & 0.47 & & 0.43, 0.39\\
    $^{150}$Nd & 0.679 & 0.589 & 0.400 & & 0.52 & 0.51, 0.52
  \end{tabular}
  \end{ruledtabular}
  \end{table}
  
Table~\ref{table:overlap} lists the values of the HFB overlap included in the
matrix ${\cal O}$. Our values agree with those of previous calculations with
similar nuclear deformation.  The overlap becomes small when the deformation of
the initial and final states are different.  That situation arises in $^{130}$Te
and $^{136}$Xe, where the initial states are spherical while the final states
are prolate and oblate, respectively.  The overlap also becomes small when the
initial or final state has no pairing gap. That is the case for neutrons in 
$^{136}$Xe.

\subsection{Contour}

To use the expression in Eq.\ \eqref{eq:M}, we must choose the contours $C_i$
and $C_f$.  We take each to be centered on the real axis and circular, with the
circle specified by the two energies $\omega_{\rm L}$ and $ \omega_{\rm R}$ at
they cross the real axis.  The radius $r$ and the center of the contour
$\omega_0$ are then given by $r = (\omega_{\rm R} - \omega_{\rm L})/2$ and
$\omega_0 =(\omega_{\rm L} + \omega_{\rm R})/2$, and every point on the contour
can be written in the form $\omega = \omega_0 + r e^{i\theta}$.  We use
$\omega_{\rm L} = 0.1$ MeV and $\omega_{\rm R}=120$ MeV for $C_i$ and
$\omega_{\rm L} = -120$ MeV and $\omega_{\rm R} = -0.1$ MeV for $C_f$ to include
all the unperturbed two-quasiparticle states within the quasiparticle-energy
cutoff.

\begin{figure}[b]
\includegraphics[width=\columnwidth]{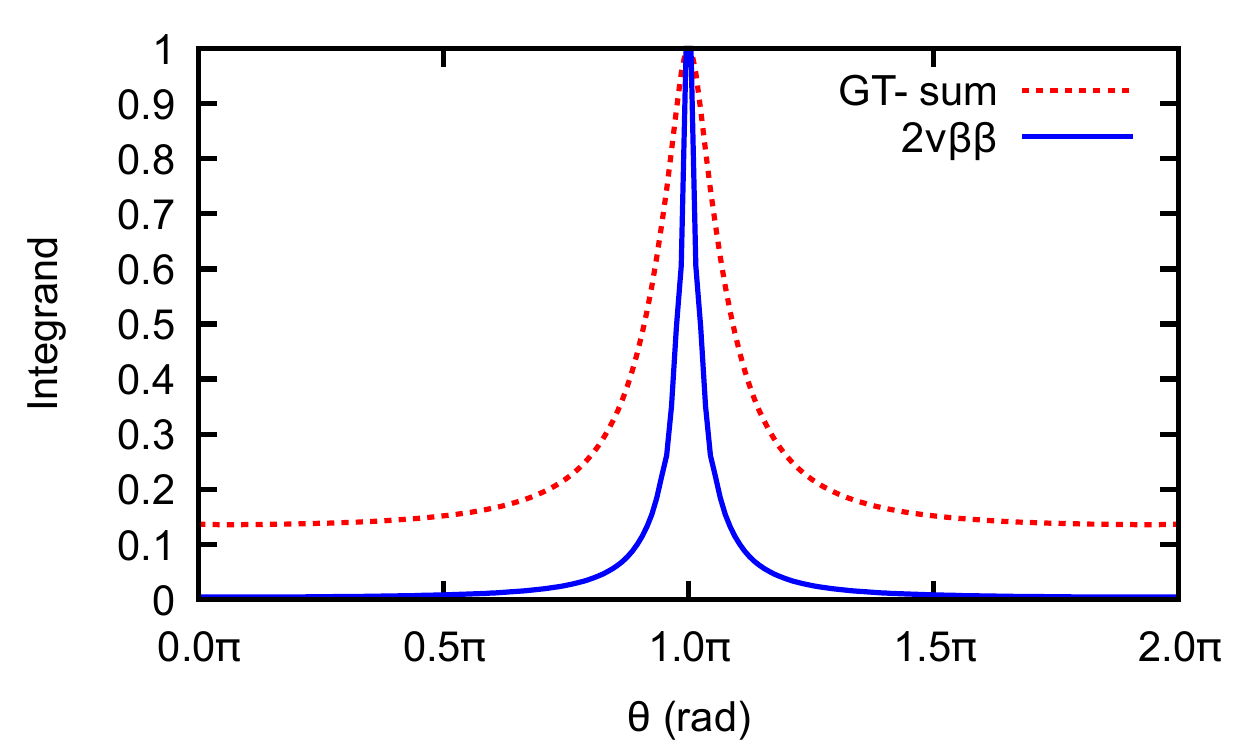}
\caption{Integrand in computation of summed Gamow-Teller strength and
$2\nu\beta\beta$ matrix element for $^{76}$Ge as a function of the angle
$\theta$ for the contour $C_i$. The integrand is normalized to 1 at
$\theta=\pi$.}
\label{fig:integrand}
\end{figure}

Figure~\ref{fig:integrand} shows the integrands for the summed GT strength and
the $2\nu\beta\beta$ matrix element in $^{76}$Ge as a function of the angle
$\theta$ for the contour $C_i$.  The main contribution to each comes from the
peak at $\theta=\pi$, the point at $\omega=\omega_{\rm L}$ where the two
contours are closest.  While the integrand of the sum is distributed broadly
along the whole contour $C_i$, the energy denominator $2/(\omega_i-\omega_f)$
concentrates the $2\nu\beta\beta$ contribution at $\theta=\pi$.  To take this
contribution into account precisely and efficiently, we introduce a parameter
$\gamma$ to control the distribution of the discretized points near the origin,
i.e., we discretize the angle $\theta$ as follows: 
\begin{align}
    \theta_k &= (m + x_k^\gamma) \pi,\\
    x_k &= -1 + 2 \frac{k-1}{n_r - 1} \quad
    (k = 1, 2, \cdots, n_r) \,, 
\end{align}
where $\gamma$ is an odd number, and $m$ is 1 for $C_i$ and $0$ for $C_f$.  The
parameter $\theta_k$ runs from 0 to $2\pi$ for $C_i$ and $-\pi$ to $\pi$ for
$C_f$.  We use $n_r=202$ and omit the contribution from $(\omega_i,\omega_f) =
(\omega_{\rm R}, -\omega_{\rm L})$, because those points are on the real axis and can be very
close to the QRPA poles, although their contribution to the nuclear matrix
element should be small because of the factor $2/(\omega_i-\omega_f)$.

Table~\ref{table:gamma} shows the dependence of the summed strengths and the
$2\nu\beta\beta$ matrix element in $^{76}$Ge on the parameter $\gamma$.  The
matrix element converges by $\gamma = 5$, which is the value we use.

\subsection{Summed Strengths}

Table~\ref{table:sumrule} shows the unweighted summed Fermi and Gamow-Teller
strengths obtained from the double contour integration for selected nuclei of
interest to experimentalists.  Integration up to 120 MeV reproduces more than
99.9\% of the Ikeda sum rule in all these nuclei. 

\begin{table}[t]
\caption{ Dependence on the discretization parameter $\gamma$ in $^{76}$Ge of
summed Fermi and Gamow-Teller strengths $m({\rm
F\pm})=\sum_{\lambda>0}|\langle\lambda,0|\hat{F}^{\rm F\pm}|0^+\rangle|^2$ and
$m({\rm GT\pm})=\sum_K(-1)^K\sum_{\lambda>0}|\langle\lambda,K|\hat{F}^{\rm
GT\pm}_K|0^+\rangle|^2$ and of the dimensionless Gamow-Teller $2\nu\beta\beta$
nuclear matrix element.  We use volume like-particle pairing and no isoscalar
pairing.
\label{table:gamma}}
\begin{ruledtabular}
\begin{tabular}{ccccc}
 $\gamma$ & \textrm{1} & \textrm{3} & \textrm{5} & \textrm{7}\\ \colrule
$m({\rm F-})$ & 12.0213 & 12.0209 & 12.0201 & 12.0189 \\ 
$m({\rm F+})$ & 0.0252 & 0.0255 & 0.0260 & 0.0269 \\ 
$m({\rm F-})-m({\rm F+})$ & 11.9961 & 11.9954 & 11.9940 & 11.9920 \\
\colrule $m({\rm GT-})$ & 37.5860 & 37.5837 & 37.5811 & 37.5774 \\
$m({\rm GT+})$ & 1.6065 & 1.6070 & 1.6085 & 1.6109 \\
$m({\rm GT-})-m({\rm GT+})$ & 35.9795 &35.9767 & 35.9726 & 35.9664 \\
\colrule $M^{2\nu} m_ec^2$ & 0.1802 & 0.1574 & 0.1574 & 0.1575 
\end{tabular}
\end{ruledtabular}
\end{table}

\begin{table}[b]
\caption{Summed Fermi and Gamow-Teller transitions, from double contour
integration ($\omega_{\rm L}=0.1$ MeV and $\omega_{\rm R}=120$ MeV,
$n_r=202$, and $\gamma=5$), as percentages of the corresponding sum rules.
\label{table:sumrule}}
\begin{ruledtabular}
\begin{tabular}{rcc}
& $\displaystyle\frac{m(\text{F}-)-m(\text{F}+)}{N-Z}$ & $\displaystyle\frac{m(\text{GT}-)-m(\text{GT}+)}{3(N-Z)}$\\ \colrule
 $^{76}$Ge & 0.9995 & 0.9992 \\
 $^{76}$Se & 0.9994 & 0.9992 \\
 $^{130}$Te & 0.9996 & 0.9993 \\
 $^{130}$Xe & 0.9996 & 0.9993 \\
 $^{136}$Xe & 0.9998 & 0.9997 \\
 $^{136}$Ba & 0.9996 & 0.9994 \\
 $^{150}$Nd & 0.9996 & 0.9994 \\
 $^{150}$Sm & 0.9996 & 0.9995 
\end{tabular}
\end{ruledtabular}
\end{table}

\subsection{\texorpdfstring{$2\nu\beta\beta$}{2nbb} matrix element}

\begin{figure}[t]
  \includegraphics[width=\columnwidth]{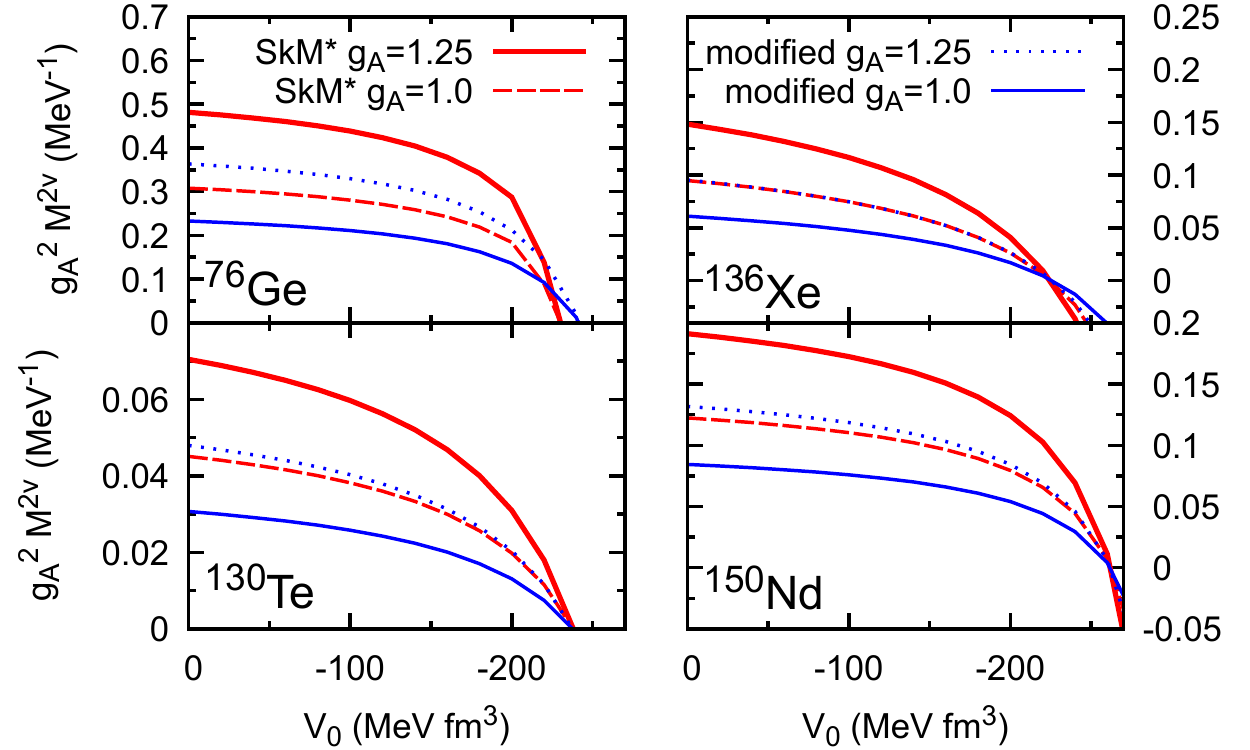}
 \caption{ Dependence on the isoscalar pairing strength of the 2$\nu\beta\beta$
 Gamow-Teller nuclear matrix element (in units of MeV$^{-1}$) for $^{76}$Ge,
 $^{130}$Te, $^{136}$Xe, and $^{150}$Nd, with the SkM* + volume pairing EDF.
 Red curves indicate the results with the time-odd functional derived from the
 SkM* interaction and blue curves the results with the modified time-odd
 functional.  The thick solid and dotted curves correspond to $g_A=1.25$, and
 the dashed and thin solid curves to the quenched value $g_A=1.0$. 
  \label{fig:2nbb_benchmark_aved3_r2sdmat}}
\end{figure}

We calculate the $2\nu\beta\beta$ matrix elements for $^{76}$Ge, $^{130}$Te,
$^{136}$Xe, and $^{150}$Nd, setting the neutron-proton isovector pairing
strength to the average of the neutron and proton like-particle pairing
strengths [$V_1=(V_n+V_p)/2$] and varying the isoscalar pairing strength $V_0$
from 0 to $-300$ MeV fm$^3$.  We use the QTDA ($\alpha=0$) to compute the
overlap among intermediate states.  Figure~\ref{fig:2nbb_benchmark_aved3_r2sdmat}
displays the dependence of the $2\nu\beta\beta$ Gamow-Teller nuclear matrix
elements on the isoscalar pairing strength.
Like the authors of that
paper, we use two values of $g_A$: one ``unquenched'' ($g_A=1.25$, though the
currently accepted value is greater than 1.27) and one quenched ($g_A=1.0$), and
compare results for the EDF SkM* with and without a modified proton-neutron
piece [$C_1^s=100$ MeV fm$^3$, $C_1^T=C_1^{\nabla s} = 0$, see
Eq.~(\ref{eq:time-odd})].  Our matrix elements agree reasonably well with those
of Ref.~\cite{PhysRevC.87.064302} in $^{130}$Te, $^{136}$Xe, and $^{150}$Nd,
while they are about twice as large in $^{76}$Ge.  

\section{global edf \label{sec:global}}

\subsection{Performance of global EDFs}

As Fig.~\ref{fig:2nbb_benchmark_aved3_r2sdmat} shows, the $2\nu\beta\beta$ and
$0\nu\beta\beta$ nuclear matrix elements are suppressed by isoscalar
proton-neutron pairing correlations that cannot be constrained from the
ground-state properties of even-even nuclei.  The usual QRPA approach uses
$2\nu\beta\beta$ decay rates to determine the strength of isoscalar pairing
separately in each decaying nucleus.  The philosophy of nuclear DFT, however, is
that one EDF parameter set should, if possible, describe all the
$\beta\beta$-decaying nuclei in the nuclear chart.  In this section we
assess the ability of globally-fit EDFs to describe $2\nu\beta\beta$ decay,
without using that observable at all in the fitting.  We use Skyrme-type EDFs, with
the isovector time-odd and isoscalar pairing parts globally fit to
single-$\beta$ decay rates and to Gamow-Teller and spin-dipole resonances.
Reference~\cite{PhysRevC.93.014304}, which deals with single-$\beta$ decay in many
isotopes, undertakes the global fitting and proposes ten parameter sets, called
1A, 1B, 1C, 1D, 1E, 2, 3A, 3B, 4, and 5, each corresponding to a different EDF.
The time-even parts for all the sets except set 2 are taken from the functional
SkO$'$ \cite{PhysRevC.60.014316}; that of parameter set 2 comes from the
functional SV-min \cite{PhysRevC.79.034310} (though tensor-density terms are
neglected, and the neutron and proton have different masses).  In both cases,
center-of-mass corrections to the mean field are neglected.  The paper uses
mixed volume-surface like-particle isovector pairing terms, fit to reproduce
odd-even staggering in ten isotopes with $50\le A \le 250$; the strengths are
$V_n = -253.75$ MeV fm$^3$, $V_p = -274.68$ MeV fm$^3$ for SkO$'$
and $V_n = -244.06$ MeV fm$^3$, $V_p = -257.90$ MeV fm$^3$ for
SV-min \cite{mustonen-priv}. 

\begin{table}[t]
\caption{ The neutron and proton pairing gaps (in MeV) and quadrupole
deformation for the lowest-energy HFB solutions in the initial and final nuclei
of the decay, computed with the SkO$'$ and SV-min EDFs, together with a mixed
pairing EDF.  Solutions with parentheses are not the lowest-energy ones, but we
use them in addition when calculating matrix elements.
\label{table:hfb-global1}}
\begin{ruledtabular}
\begin{tabular}{ccccccc}
&& \textrm{SkO$'$} & & & \textrm{SV-min} & \\
 & $\Delta_n$ & $\Delta_p$ & $\beta$ & $\Delta_n$ & $\Delta_p$ & $\beta$ \\
 \colrule $^{48}$Ca & 0.771& 0.000& 0.000& 0.793& 0.000& 0.000 \\ 
$^{48}$Ti & 1.270& 1.386& 0.000& 1.275& 1.309& 0.000 \\ \colrule
$^{76}$Ge & 1.063& 1.189& 0.136& 1.123& 1.094& 0.131 \\
$^{76}$Se & 1.134& 1.532& 0.000& 1.165& 1.352& 0.000 \\ \colrule
$^{82}$Se & 0.619& 1.106& 0.152& 0.689& 1.124& 0.134 \\
$^{82}$Kr & 1.014& 1.353& 0.112& 1.041& 1.230& 0.101 \\ \colrule
$^{96}$Zr & 1.153& 1.133& $-$0.173& 1.041& 0.986& 0.000 \\
           & (1.354 & 1.129 &0.000)& \\
$^{96}$Mo & 1.202& 1.174& 0.000& 0.991& 1.090& 0.000 \\ \colrule
$^{100}$Mo & 1.200& 1.089& $-$0.192& 1.299& 1.078& 0.000 \\
           & (1.123& 1.246& 0.214)& \\
$^{100}$Ru & 0.994& 1.092& 0.186& 1.189& 1.137& 0.000 \\ \colrule
$^{116}$Cd & 1.430& 0.854& 0.000& 1.463& 0.492& 0.120 \\
$^{116}$Sn & 1.406& 0.000& 0.000& 1.553& 0.000& 0.000 \\ \colrule
$^{128}$Te & 1.139& 0.970& 0.000& 1.209& 0.907& 0.000 \\
$^{128}$Xe & 1.136& 0.912& 0.142& 1.152& 0.841& 0.156 \\ 
           &(1.147& 1.064&$-$0.112)&(1.179& 0.986&$-$0.122) \\ \colrule
$^{130}$Te & 1.013& 0.971& 0.000& 1.043& 0.902& 0.000 \\
$^{130}$Xe & 1.051& 1.001& 0.111& 1.077& 0.947& 0.118 \\ \colrule
$^{136}$Xe & 0.000& 1.180& 0.000& 0.000& 1.143& 0.000 \\
$^{136}$Ba & 0.767& 1.349& 0.000& 0.775& 1.296& 0.000 \\ \colrule
$^{150}$Nd & 0.962& 0.686& 0.311& 0.886& 0.830& 0.266 \\
$^{150}$Sm & 0.901& 1.074& 0.238& 0.823& 1.101& 0.203 \\ \colrule
$^{238}$U & 0.863& 0.735& 0.265& 0.763& 0.596& 0.269 \\
$^{238}$Pu & 0.828& 0.640& 0.269& 0.745& 0.572& 0.272
\end{tabular}
\end{ruledtabular}
\end{table}

The isovector time-odd part of any Skyrme-type EDF is given by
\begin{align}
\chi^{\rm odd}_1(\bm{r}) &= 
C_1^s[\rho_0]\bm{s}_1^2 + C_1^{\Delta s} \bm{s}_1\cdot \Delta \bm{s}_1 + C_1^j \bm{j}_1^2
\nonumber \\
&\quad + C_1^T \bm{s}_1\cdot \bm{T}_1 
+ C_1^{\nabla j} \bm{s}_1\cdot \bm{\nabla}\times \bm{j}_1
\nonumber \\ &\quad 
+ C_1^F \bm{s}_1\cdot \bm{F}_1 
+ C_1^{\nabla s} ( \bm{\nabla}\cdot\bm{s}_1)^2 \,,
\label{eq:time-odd}
\end{align}
where $\bm{s}_1, \bm{j}_1, \bm{T}_1$, and $\bm{F}_1$ are the isovector spin,
current, spin-kinetic, and tensor-kinetic densities, respectively.  The
isoscalar pairing functional in all these parametrizations has the mixed
density dependence
\begin{align}
\tilde{\chi}_0(\bm{r}) = \frac{V_0}{4}\left[
1 - \frac{1}{2} \frac{\rho_0(\bm{r})}{\rho_c} 
\right] | \tilde{\bm{s}}_0(\bm{r})|^2 \,,
\end{align}
where $\tilde{\bm{s}}_0$ is the isoscalar pair density, $\rho_c=0.16$ fm$^{-3}$,
and $\rho_0$ is the usual isoscalar density.  In the parameter sets 1A, 1B, 1C,
1D, 1E, only $C_1^s$ (with no density dependence) and $V_0$ are fit.  In sets 3A
and 3B $C_1^T$ and $C_1^F$ are fit as well.  In the parameter set 4, $C_1^j,
C_1^{\nabla j}$, and $C_1^{\nabla s}$ are adjusted, while other parameters are
the same as in set 3A.  In set 5, $V_0$, $C_1^s$, and $C_1^j$ are fit.  

\begin{table}[t]
\caption{The HFB overlap $\langle 0^+_{f,{\rm HFB}}|0^+_{{\rm HFB},i}\rangle$
between the initial and the final states of the double-beta decay, computed with
SkO$'$ and SV-min EDFs.
The numbers in parentheses denote powers of 10. \label{table:overlap-global1}}
\begin{ruledtabular} 
  \begin{tabular}{cccccccccc}
     & & SkO$'$ & & & SV-min & & \\
     & neutron & proton & total & neutron & proton & total & \\ \hline
    $^{48}$Ca & 0.764 & 0.513 & 0.392 & 0.776 & 0.512 & 0.398 \\
    $^{76}$Ge & 0.577 & 0.559 & 0.323 & 0.586 & 0.587 & 0.344 \\
    $^{82}$Se & 0.729 & 0.829 & 0.604 & 0.772 & 0.862 & 0.665 \\
    $^{96}$Zr & 0.283 & 0.306 & 0.087 & 0.882 & 0.877 & 0.774 \\
    (sph.$\to$sph.) & 0.915 & 0.893 & 0.818 & \\
    $^{100}$Mo & $1.8(-3)$& $1.4(-2)$ & $2.6(-5)$ &0.914 & 0.905 & 0.828 \\
    (pro.$\to$pro.) & 0.864 & 0.875 & 0.755 \\
    $^{116}$Cd & 0.932 & 0.521 & 0.485 & 0.507 & 0.293 & 0.148 \\
    $^{128}$Te & 0.342 & 0.388 & 0.133 & 0.294 & 0.343 & 0.101 \\
    (obl.$^{128}$Xe) & 0.440 & 0.533 & 0.235 & 0.403 & 0.487 & 0.197 \\
    $^{130}$Te & 0.489 & 0.523 & 0.256 & 0.464 & 0.509 & 0.236 \\
    $^{136}$Xe & 0.517 & 0.921 & 0.476 & 0.522 & 0.931 & 0.486 \\
    $^{150}$Nd & 0.624 & 0.601 & 0.375 & 0.711 & 0.683 & 0.485 \\
    $^{238}$U & 0.912 & 0.882 & 0.805 & 0.902 & 0.873 & 0.787
  \end{tabular}
  \end{ruledtabular}
  \end{table}

Table~\ref{table:hfb-global1} lists the pairing gaps and quadrupole deformation
of the HFB states used to compute $2\nu\beta\beta$ nuclear matrix elements.
Neutron pairing collapses only in $^{136}$Xe and proton pairing collapses in
$^{48}$Ca and $^{116}$Sn.
SkO$'$ and SV-min cause different
amounts of deformation.  $^{96}$Zr, $^{100}$Mo, and $^{100}$Ru are oblate,
oblate, and prolate (respectively) with SkO$'$, while they are all spherical
with SV-min.  $^{116}$Cd is spherical with SkO$'$, but is
prolate with SV-min.
 
Table \ref{table:overlap-global1} contains the overlaps of the initial and final
HFB vacua.  Significant differences in deformation and pairing between the two
HFB states lead to small overlaps, and because the two EDFs can produce
different levels of deformation and pairing in any nucleus, the overlaps depend
significantly on the EDF.  In $^{96}$Zr and $^{100}$Mo, the HFB overlaps with
SkO$'$ are extremely small because the initial state is oblate and the final state
spherical or prolate.  In $^{116}$Cd, the HFB overlap with
SV-min is smaller for a similar reason.  The QRPA may not be adequate when the
overlaps, like those with SkO$'$ in $^{100}$Mo, are very small.  Our treatment
omits both projection onto states with good angular momentum, which involves the mixing of states with different orientations, and the fluctuation in shape and pairing
captured, e.g., by the generator coordinate method \cite{LopezVaquero2011520,PhysRevC.90.031301}.
The effects of the physics we have neglected can be significant when the matrix
elements are small at the HFB or QRPA levels.

\begin{figure}[t]
\includegraphics[width=\columnwidth]{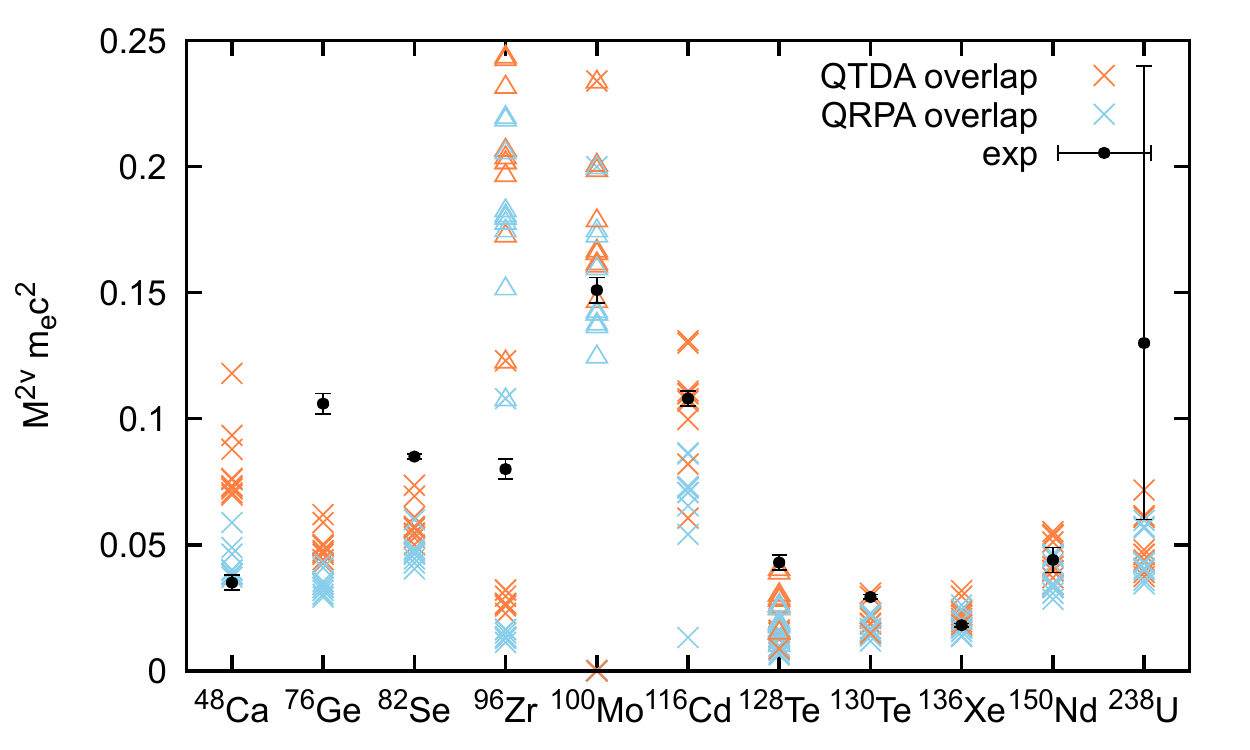}
\caption{Comparison of dimensionless $2\nu\beta\beta$ nuclear matrix elements obtained from
global EDFs with experimental values.  The matrix elements computed with the
lowest-energy HFB solutions are marked with crosses, while those elements
computed with the other HFB solutions are marked with triangles.  Orange symbols
come from computations with the QTDA overlap and blue symbols from computations
with the QRPA overlap.  The EDFs that give rise to each particular point appear
in Table \ref{table:2nbb-global1} in Appendix \ref{sec:appen-num}.
\label{fig:2nbb}}
\end{figure}

\begin{figure*}[t]
\includegraphics[width=\textwidth]{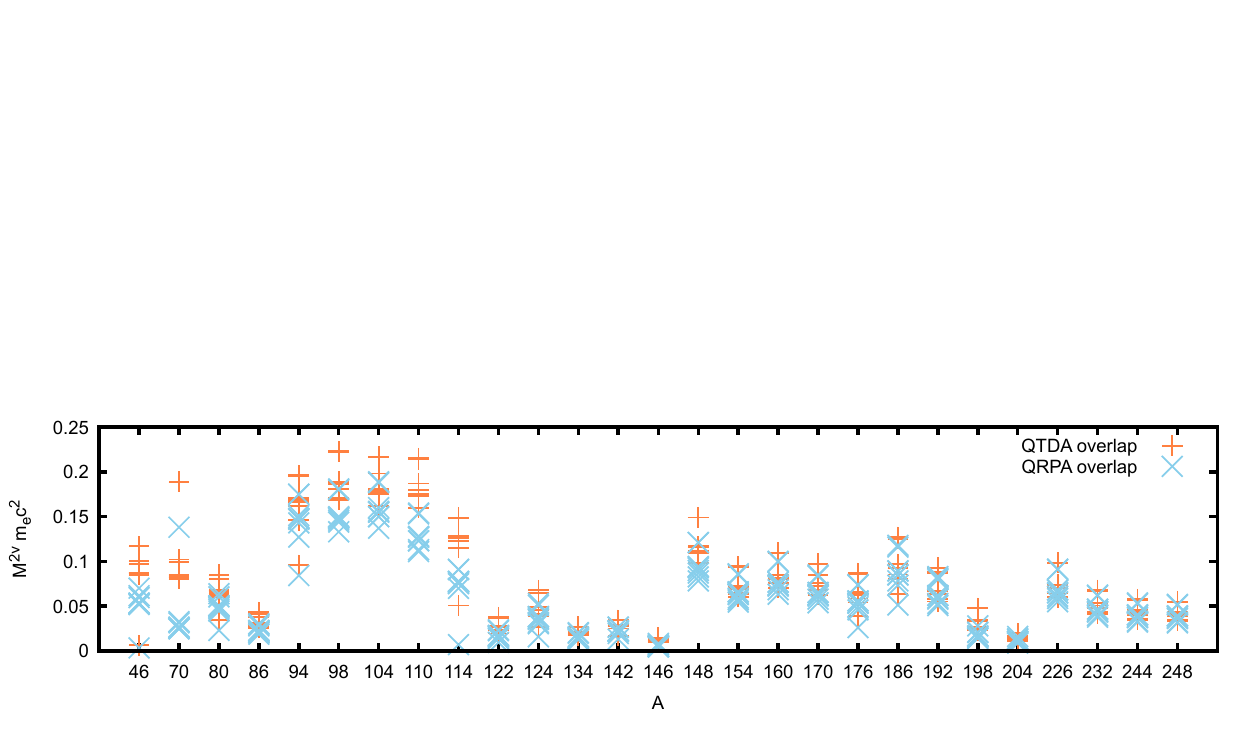}
\caption{Dimensionless $2\nu\beta\beta$ nuclear matrix elements calculated with the global
EDFs.  The matrix elements from parameter set 4 are excluded.
\label{fig:2nbb_prediction}}
\end{figure*}

In Fig.~\ref{fig:2nbb} we compare the Gamow-Teller $2\nu\beta\beta$ nuclear
matrix element, scaled by $g_A^2 m_e c^2$ to be dimensionless, that results from
calculations with the ten different SkO$'$- and SV-min-based EDFs discussed just above.  We
also show the experimental matrix elements, extracted from the half-lives in
Ref.~\cite{universe.6.159}. We use a quenched axial-vector coupling constant
$g_A=1.0$ to match the value from Ref.~\cite{PhysRevC.93.014304}, which
determines the EDF parameters.  Despite the differences among the EDFs in the
pieces of the functional that were fit and in the data chosen to fit them, the
$2\nu\beta\beta$ matrix elements that they produce are quite close to one
another in some of the heavier nuclei.  This fact means that the parts of the
EDF that affect the $2\nu\beta\beta$ matrix element are determined almost fully
by the $\beta$ decay rates and giant resonance energies used in fitting them.
In some lighter isotopes such as $^{48}$Ca, $^{96}$Zr, and $^{100}$Mo, on the
other hand, the values of the nuclear matrix element, like those of the overlap,
depend significantly on the EDF.  Although such matrix elements may provide an
additional constraint on the pnEDF, the disagreement also suggests, as we noted
earlier, that correlations that escape the QRPA are important
\cite{Rodriguez2011436,0954-3899-44-3-034002}. Table~\ref{table:2nbb-global1}
in Appendix \ref{sec:appen-num} contains more detail than Fig.\ \ref{fig:2nbb},
in particular the values for each individual EDF of all the matrix elements.

In some nuclei, such as $^{76}$Ge, the EDFs all produce comparable values for
the matrix element, but those values are quite different from the experimental
one.  The reason for the discrepancy, again, is the quite different degrees of
deformation in the initial and final nuclei, a difference that in reality is
probably made less significant by shape fluctuations.  In other nuclei,
$^{96}$Zr, $^{100}$Mo, and $^{128}$Te (and especially the first two), the values
span a wide range.  The reason is that two local minima appear in the initial
isotopes, and the value of the $2\nu\beta\beta$ matrix element depends strongly
on which minimum is used. 
In these two cases, the HFB overlaps associated with the lowest minima for SkO$'$ are very small (0.087 and 2.6$\times$10$^{-5}$ in $^{96}$Zr and $^{100}$Mo), and the $2\nu\beta\beta$ matrix elements from the lowest minimum are consequently smaller than the experimental values, 
while the matrix elements associated with the other HFB solutions are larger than or comparable to the experimental values.
Correlations that admix states near those other minima, if they were taken into account, would probably increase 
the $2\nu\beta\beta$ matrix elements produced by the lowest minima. Such admixtures are beyond what the QRPA includes, however, and seeing their effects would require an approximation such as the generator-coordinate method.

Figure \ref{fig:2nbb} also shows that overlaps computed with the
QTDA prescription result in larger matrix elements than those computed with the
QRPA prescription. 

\subsection{Predictions}

Using the same global EDFs as in the previous section, we compute the
$2\nu\beta\beta$ matrix elements for all the nuclei in which that decay might
conceivably be observed: $^{46}$Ca, $^{70}$Zn, $^{80}$Se, $^{86}$Kr, $^{94}$Zr,
$^{98}$Mo, $^{104}$Ru, $^{110}$Pd, $^{114}$Cd, $^{122}$Sn, $^{124}$Sn, $^{134}$Xe, $^{142}$Ce,
$^{146}$Nd, $^{148}$Nd, $^{154}$Sm, $^{160}$Gd, $^{170}$Er, $^{176}$Yb,
$^{186}$W, $^{192}$Os, $^{198}$Pt, $^{204}$Hg, $^{226}$Ra, $^{232}$Th,
$^{244}$Pu, and $^{248}$Cm.  Figure \ref{fig:2nbb_prediction}
summarizes the results, while Table \ref{table:2nbb-prediction} in Appendix
\ref{sec:appen-num} indicates the individual EDFs responsible for each symbol in
the figure. 
We emphasize that we are able to make these predictions only because we use EDFs
that are fit globally and without considering $2\nu\beta\beta$ half-lives.  In
typical QRPA calculations, by contrast, the strength of isoscalar pairing is
adjusted in each nucleus individually to reproduce the $2\nu\beta\beta$ half-life.  

As we mentioned in the previous section, the QRPA nuclear matrix elements may
not be reliable if the deformations of the initial and final states of the decay
are different.  The deformation parameters differ by more than 0.1 for the following
decays: $^{70}$Zn $\to$ $^{70}$Ge, $^{80}$Se $\to$ $^{80}$Kr, $^{134}$Xe $\to$
$^{134}$Ba, and $^{146}$Nd $\to$ $^{146}$Sm with SkO$'$ and $^{80}$Se $\to$
$^{80}$Kr, $^{114}$Cd $\to$ $^{114}$Sn, $^{122}$Sn $\to$ $^{122}$Te, $^{134}$Xe
$\to$ $^{134}$Ba, and $^{146}$Nd $\to$ $^{146}$Sm with SV-min.  We also saw
earlier that the QRPA can go awry if shape mixing is important.  A full
treatment of shape mixing requires something like the generator coordinate
method \cite{Ring-Schuck,RevModPhys.75.121}, but we can get a good idea of when
it will be significant by examining potential energy curves.  These turn out to
be broad near the minimum for the nuclei $^{46}$Ti, $^{70}$Ge, $^{94}$Zr (only
SkO$'$), $^{98}$Mo, $^{104}$Ru, $^{110}$Pd, $^{114}$Cd, $^{122}$Te,
$^{124}$Te, $^{134}$Ba,
$^{142}$Ce, $^{198}$Pt, and $^{198}$Hg.  Unfortunately, the generator coordinate
method, while it has been applied to $0\nu\beta\beta$ decay
\cite{PhysRevC.90.031301,PhysRevC.93.014305,PhysRevLett.105.252503,Rodriguez2011436,Rodriguez2013174,PhysRevLett.124.232501,PhysRevC.98.054311,PhysRevC.95.024305,PhysRevC.91.024316,PhysRevC.90.054309,PhysRevC.96.054310,
PhysRevC.98.064324, PhysRevC.100.031303} is difficult to apply to
$2\nu\beta\beta$ decay because the closure approximation is poor there and a
complete set of intermediate states is required. 

With the parameter set 4, the pnFAM converges more slowly than with the other
parameter sets, and the resulting matrix elements are often quite different from
those produced by the other sets.  Thus we exclude set 4 from the distribution
of the nuclear matrix elements shown in Fig.~\ref{fig:2nbb_prediction}.  We see
better agreement among the other EDFs in heavier isotopes as a general rule, and
the QTDA prescription for the overlap again leads to larger numbers than does
QRPA prescription.

\section{Conclusions \label{sec:conclusion}}

We have presented a computationally efficient framework for calculating the
matrix elements for two-neutrino double-beta decay within nuclear
density functional theory.  We employ the finite amplitude method to compute the
QRPA approximation to the matrix elements.  Our approach allows large
single-particle model spaces and the use of a single nuclear EDF for all nuclei.
It also eliminates the need to truncate two-quasiparticle spaces.

We first used harmonic-oscillator-based HFB and FAM codes together with familiar
EDFs to compute the $2\nu\beta\beta$ matrix elements in a few important nuclei,
comparing the results with those obtained previously by diagonalizing the QRPA
matrix.  Using EDFs that had been fit globally to single-$\beta$ decay rates and
giant-resonance energies, we then computed the $2\nu\beta\beta$ matrix elements
in all nuclei in which double-beta decay has or could be observed.  Agreement
with the matrix elements extracted from already measured half-lives is good in
general, and we offered predictions for those nuclei that have unmeasured
half-lives.  

Although we focus on $2\nu\beta\beta$ decay in this paper, we can also compute
double-electron capture matrix elements in the same way.  The most interesting
extension of our work is to neutrinoless double-beta decay.  The presence of a
neutrino propagator in that matrix element, however, will make that process more
challenging to treat than $2\nu\beta\beta$ decay. 

\section*{Acknowledgments}

We are grateful to Mika T. Mustonen for the valuable discussions.  This work is
supported by JSPS KAKENHI Grants No.~17H05194, No.~19KK0343, No.~20H05242,
and No.~20K03964, and by the U.S. Department of Energy, Office of Science, Office
of Nuclear Physics, under grant No. DE-FG02-97ER41019.  
This research was conducted in part during the INT program INT-17-2a ``Neutrinoless Double-Beta Decay," at the Institute for Nuclear Theory, University of Washington.
Numerical calculations were performed at the Oakforst-PACS Systems through the Multidisciplinary Cooperative Research Program of the Center for Computational Sciences,
University of Tsukuba.

\appendix

\section{Overlap \label{sec:overlap}}

\subsection{QRPA overlap}

We follow the discussion in Ref.~\cite{Simkovic2004321} to evaluate the overlap
$\langle \lambda_f,K|\lambda_i,K\rangle $ of two intermediate states.  The QRPA
phonon operators that excite the initial HFB state are related to the those that
excite the final state by
\begin{align}
  \hat{\cal Q}^{\lambda_i\dag}_{K} = \sum_{\lambda_f} \left(
  a_{\lambda_i\lambda_f} \hat{\cal Q}^{\lambda_f\dag}_{K} +
  b_{\lambda_i\lambda_f}\hat{\tilde{{\cal Q}}}^{\lambda_f}_{K} \right) \,,
\end{align}
where $\hat{\tilde{{\cal Q}}}_K^{\lambda} = \hat{{\cal Q}}_{-K}^\lambda$.  This
relation is based on the fact that both operators span the complete set of
two-quasiparticle states with angular momentum projection $K$.  The overlap of
the intermediate state can be written in terms of the phonon operators as 
\begin{align}
\label{eq:ovl}
  \langle \lambda_f,K|\lambda_i,K\rangle &= \langle 0^+_{f,{\rm QRPA}}|\hat{\cal Q}^{\lambda_f}_K\hat{\cal Q}^{\lambda_i\dag}_K|0^+_{i,{\rm QRPA}}\rangle\nonumber\\
  &= \sum_{\lambda_f'} \left(
    \langle 0^+_{f,{\rm QRPA}}|\hat{\cal Q}^{\lambda_f}_K \hat{\cal Q}^{\lambda_f'\dag}_K|0^+_{i,{\rm QRPA}}\rangle a_{\lambda_i \lambda_f'}
    \right. \nonumber \\
    &\quad + \left.
    \langle 0^+_{f,{\rm QRPA}}|\hat{\cal Q}^{\lambda_f}_K \hat{\cal Q}^{\lambda_f'}_{-K}|0^+_{i,{\rm QRPA}}\rangle b_{\lambda_i \lambda_f'}
    \right) \nonumber \\
  &\approx a_{\lambda_i\lambda_f} \langle 0^+_{f,{\rm HFB}}| 0^+_{i,{\rm
  HFB}}\rangle \,,
\end{align}
where we neglect the term proportional to $b_{\lambda_i\lambda_f'}$, because it
involves a two-phonon state, and approximate the overlap between the two QRPA
correlated ground states. 

We have two sets of the quasiparticles, one defined for the initial HFB state
and the other for the final state:
\begin{align}
  \hat{a}^{(i)}_{\mu}|0^+_{i,{\rm HFB}}\rangle = 0, \quad
  \hat{a}^{(f)}_{\mu}|0^+_{f,{\rm HFB}}\rangle = 0,
\end{align}
with $\mu$ a proton or neutron single-particle state with positive angular
momentum $j_z$ along the symmetry axis.  We write the transformation between
the two sets of the quasiparticles in the form 
\begin{align}
  \hat{a}^{(i)\dag}_{\mu} &= {\sum_{\nu\in\tau}}' \left(
  {\cal R}_{\mu\nu} \hat{a}^{(f)\dag}_{\nu} + {\cal S}_{\mu\bar{\nu}} \hat{a}^{(f)}_{\bar{\nu}}
  \right), \\
  \hat{a}^{(i)\dag}_{\bar{\mu}} &= {\sum_{\nu\in\tau}}' \left(
  {\cal R}_{\bar{\mu}\bar{\nu}} \hat{a}^{(f)\dag}_{\bar{\nu}} + {\cal S}_{\bar{\mu}\nu} \hat{a}^{(f)}_{\nu}
  \right),
\end{align}
where ${\sum}'$ means that the summation is only over states with $j_z > 0$,
and the notation $\nu \in \tau$ means that index $\nu$ corresponds to the same
kind of particle (proton or neutron) as does the index $\mu$ on the left side of
the equation.

The relation
\begin{align}
    {\cal R}^T {\cal R}^\ast + {\cal S}^\dag {\cal S} &= I, \\
    {\cal R}^T {\cal S}^\ast + {\cal S}^\dag {\cal R} &= 0
\end{align}
follows from the unitarity of the transformation.

This transformation is defined in the full quasiparticle model space; any
quasiparticle cutoff thus breaks unitarity.  Because the matrix composed of
${\cal R}$ and ${\cal S}$ is also unitary, the inverse transformation is given
by
\begin{align}
 \hat{a}^{(f)\dag}_{\mu} &= {\sum_{\nu\in\tau}}' \left(
  {\cal S}_{\bar{\nu}\mu} \hat{a}^{(i)}_{\bar{\nu}} + {\cal R}^{\ast}_{\nu\mu} \hat{a}^{(i)\dag}_{\nu}
  \right), \\
  \hat{a}^{(f)\dag}_{\bar{\mu}} &= {\sum_{\nu\in\tau}}' \left(
  {\cal S}_{\nu\bar{\mu}} \hat{a}^{(i)}_{\nu} + {\cal R}^{\ast}_{\bar{\nu}\bar{\mu}} \hat{a}^{(i)\dag}_{\bar{\nu}}
  \right).
\end{align}
Using the Bogoliubov transformation
\begin{align}
\hat{a}^{(i/f)\dag}_{\mu} &= {\sum_{k\in\tau}}' V^{(i/f)}_{\bar{k}\mu} \hat{c}_{\bar{k}} + U^{(i/f)}_{k\mu} \hat{c}^\dag_{k}, \\
 \hat{a}^{(i/f)\dag}_{\bar{\mu}} &= {\sum_{k\in\tau}}' V^{(i/f)}_{k\bar{\mu}} \hat{c}_{k} + U^{(i/f)}_{\bar{k}\bar{\mu}} \hat{c}^\dag_{\bar{k}},
\end{align}
we can write the matrix elements of ${\cal R}$ and ${\cal S}$ in the form 
\begin{align}
  {\cal R}_{\mu\nu} &= {\sum_{k\in\tau}}' V^{(i)}_{\bar{k}\mu} V^{(f)\ast}_{\bar{k}\nu} + U^{(i)}_{k\mu} U^{(f)\ast}_{k\nu}, \\ 
  {\cal R}_{\bar{\mu}\bar{\nu}} &= {\sum_{k\in\tau}}' V^{(i)}_{k\bar{\mu}} V^{(f)\ast}_{k\bar{\nu}} + U^{(i)}_{\bar{k}\bar{\mu}} U^{(f)\ast}_{\bar{k}\bar{\nu}}, \\
  {\cal S}_{\mu\bar{\nu}} &= {\sum_{k\in\tau}}' V^{(i)}_{\bar{k}\mu} U^{(f)}_{\bar{k}\bar{\nu}} + U^{(i)}_{k\mu} V^{(f)}_{k\bar{\nu}}, \\ 
  {\cal S}_{\bar{\mu}\nu} &= {\sum_{k\in\tau}}' V^{(i)}_{k\bar{\mu}} U^{(f)}_{k\nu} + U^{(i)}_{\bar{k}\bar{\mu}} V^{(f)}_{\bar{k}\nu}.
\end{align}
Defining the proton-neutron two-quasiparticle creation and annihilation
operators
\begin{align}
  \bm{A}^{(i)\dag}(pn,K) \equiv \hat{a}^{(i)\dag}_{p}\hat{a}^{(i)\dag}_{n}, 
  \quad
  \bm{A}^{(i)\dag}(\bar{p}\bar{n},K) \equiv  \hat{a}^{(i)\dag}_{\bar{p}}\hat{a}^{(i)\dag}_{\bar{n}}, 
\end{align}
we can relate the two-quasiparticle operators defined with respect to the
initial and final HFB states in the following way:
\begin{align}
  \bm{A}^{(i)\dag}(pn,K) &= {\sum_{p'n'}}'\left[
    {\cal R}_{pp'}{\cal R}_{nn'}\bm{A}^{(f)\dag}(p'n',K)
    \right.\nonumber \\ &\quad \left.
    -{\cal S}_{p\bar{p}'}{\cal S}_{n\bar{n}'}\bm{A}^{(f)}(\bar{p}'\bar{n}',K)   
    \right] + (\hat{a}^\dag\hat{a}{\rm -terms}), \\
  \bm{A}^{(i)}(\bar{p}\bar{n},K) &= {\sum_{p'n'}}'\left[
    {\cal R}_{\bar{p}\bar{p}'}^{\ast}{\cal R}_{\bar{n}\bar{n}'}^{\ast}\bm{A}^{(f)}(\bar{p}'\bar{n}',K) 
        \right.\nonumber \\ &\quad \left.
    -
    {\cal S}_{\bar{p}p'}^{\ast}{\cal S}_{\bar{n}n'}^{\ast}\bm{A}^{(f)\dag}(p'n',K)   
    \right]+ (\hat{a}^\dag\hat{a}{\rm -terms}).
\end{align}

The QRPA phonon operator is a combination of two-quasiparticle creation and
annihilation operators:  
\begin{align}
  \hat{\cal Q}^{\lambda_i\dag}_K &= {\sum_{pn}}' \left[ X^{\lambda_i}_{pn,K} \bm{A}^{(i)\dag}(pn,K) - Y^{\lambda_i}_{pn,K} \bm{A}^{(i)}(\bar{p}\bar{n},K) \right], \\
   \hat{\tilde{{\cal Q}}}^{\lambda_i}_K &= {\sum_{pn}}' \left[X^{\lambda_i}_{pn,K} \bm{A}^{(i)}(\bar{p}\bar{n},K) - Y^{\lambda_i}_{pn,K} \bm{A}^{(i)\dag}(pn,K) \right].
\end{align}
Inverting this yields the relation 
\begin{align}
  \bm{A}^{(f)\dag}(pn,K) &= \sum_{\lambda_f} \left[
    X^{\lambda_f\ast}_{pn,K} \hat{\cal Q}^{\lambda_f\dag}_{K} + Y^{\lambda_f\ast}_{pn,K} \hat{\tilde{{\cal Q}}}^{\lambda_f}_{K}    
    \right],\\
  \bm{A}^{(f)}(\bar{p}\bar{n},K) &= \sum_{\lambda_f} \left[
    X^{\lambda_f\ast}_{pn,K} \hat{\tilde{{\cal Q}}}^{\lambda_f}_{K} + Y^{\lambda_f\ast}_{pn,K} \hat{\cal Q}^{\lambda_f\dag}_{K}    
    \right],  
\end{align}
which leads to an expression for the $a$ matrix in Eq.\ \eqref{eq:ovl}:
\begin{align}
  a_{\lambda_i\lambda_f} &= {\sum_{pnp'n'}}' \left[
     X^{\lambda_f\ast}_{p'n',K} {\cal R}_{pp'}{\cal R}_{nn'} X^{\lambda_i}_{pn,K}
     \right.\nonumber \\ &\quad \left.
  -  Y^{\lambda_f\ast}_{p'n',K} {\cal R}_{\bar{p}\bar{p}'}^{\ast}{\cal R}_{\bar{n}\bar{n}'}^{\ast} Y^{\lambda_i}_{pn,K}
       \right.\nonumber \\ &\quad \left.
  +  X^{\lambda_f\ast}_{p'n',K} {\cal S}_{\bar{p}p'}^{\ast}{\cal S}_{\bar{n}n'}^{\ast} Y^{\lambda_i}_{pn,K}
      \right.\nonumber \\ &\quad \left.
  -  Y^{\lambda_f\ast}_{p'n',K} {\cal S}_{p\bar{p}'}{\cal S}_{n\bar{n}'} X^{\lambda_i}_{pn,K}
  \right]. \label{eq:a}
\end{align}
In Ref.~\cite{Simkovic2004321}, the contribution from ${\cal S}$ in
Eq.~(\ref{eq:a}) is neglected. 

The overlap between the HFB states is given by the Onishi formula, e.g.,  in
Eq.\ (E.49) of Ref.\ \cite{Ring-Schuck}:  
\begin{align}
{\cal N}^{-1} &= 
    \langle 0^+_{f,{\rm HFB}}|0^+_{i,{\rm HFB}}\rangle
    =
    \langle 0^+_{i,{\rm HFB}}|0^+_{f,{\rm HFB}}\rangle
    \nonumber \\ &
    =  \left(
        \det {\cal R}^T  \right)^{\frac{1}{2}}
    = \left[ \det\left( 1 + {\cal D}^\dag {\cal D}\right) \right]^{-\frac{1}{4}}, \label{eq:overlap}
\end{align}
where ${\cal D}$ is a skew-symmetric matrix that determines the relation between
the initial and final HFB states through
\begin{align}
  |0^+_{i,{\rm HFB}}\rangle 
  &= {\cal N}^{-1} \exp\left( \sum_\tau \sum_{\mu\nu\in\tau} {\cal D}_{\mu\nu}\hat{a}^{(f)\dag}_\mu \hat{a}^{(f)\dag}_\nu \right) 
  |0^+_{f,{\rm HFB}}\rangle, \label{eq:Thouless}
\end{align}
and satisfies the relation
\begin{align}
{\cal D}= {\cal S}^\dag({\cal R}^\dag)^{-1}
= - ({\cal R}^{-1} {\cal S})^\ast 
= -{\cal D}^T. \label{eq:D}
\end{align}
Thus we end up with 
\begin{align}
\label{eq:qrpa-ovl}
  \langle \lambda_f,K|\lambda_i,K\rangle &= (\det {\cal R})^{\frac{1}{2}}
   {\sum_{pnp'n'}}' {\cal R}_{pp'} {\cal R}_{nn'}
    \nonumber \\
  &\quad\times  \left(X^{\lambda_f\ast}_{p'n',K} X^{\lambda_i}_{pn,K} - Y^{\lambda_f\ast}_{p'n',K} Y^{\lambda_i}_{pn,K}\right).
\end{align}

\subsection{QTDA overlap}

Reference~\cite{PhysRevC.87.064302} uses the QTDA to evaluate the overlap among
intermediate states, which are given by 
\begin{align}
    |\lambda_{i/f},K \rangle = \sum_{pn}
    X_{\mu\nu,K}^{\lambda_{i/f}} \hat{a}^{(i/f)\dag}_{p} \hat{a}^{(i/f)\dag}_{n} |0^+_{i/f,{\rm HFB}}\rangle. \label{eq:QTDA}
\end{align}
From Eqs.~(\ref{eq:Thouless}) and (\ref{eq:QTDA}), we find that
\begin{align}
\label{eq:qtda-ovl}
  \langle \lambda_f,K| \lambda_i,K\rangle 
&=
( \det{\cal R})^{\frac{1}{2}}
{\sum_{pnp'n'}}'
X_{p'n',K}^{\lambda_f\ast} X_{pn,K}^{\lambda_i} \nonumber \\
& \quad \times 
\left( {\cal R}_{pp'} + 2\sum_{p''} {\cal S}_{pp''} {\cal D}_{p''p'} \right)
\nonumber \\ & \quad \times
\left( {\cal R}_{nn'} + 2\sum_{n''} {\cal S}_{nn''} {\cal D}_{n''n'} \right).
\end{align}
The two QRPA and QTDA overlaps in Eqs.\ \eqref{eq:qrpa-ovl} and
\eqref{eq:qtda-ovl} can be written in the same form as
\begin{align}
  \langle \lambda_f,K|\lambda_i,K\rangle &= 
  {\sum_{pnp'n'}}' \left(X^{\lambda_f\ast}_{p'n',K} X^{\lambda_i}_{pn,K} - \alpha Y^{\lambda_f\ast}_{p'n',K} Y^{\lambda_i}_{pn,K}\right) 
  \nonumber \\ &\quad \times 
  {\cal O}_{pp'}(\alpha) {\cal O}_{nn'}(\alpha),
\end{align}
where ${\cal O}$ is a matrix that does not depend on the QRPA and includes the
HFB overlap and the transformation relating the initial and final quasiparticle
states:
\begin{align}
  {\cal O}_{\rho\rho'}(\alpha) = \left[\det {\cal R}^{(\tau)} \right]^{\frac{1}{2}}
  \left[ {\cal R}_{\rho\rho'} + 2(1-\alpha) {\sum_{\rho''\in\tau}}'  {\cal S}_{\rho\rho''}
  {\cal D}_{\rho''\rho'}
  \right].
\end{align}
Here $\rho, \rho'$ are both proton or both neutron states, and ${\cal
R}^{(\tau)}$ is the neutron or proton part of the matrix ${\cal R}$.  The QRPA
expression in Ref.~\cite{Simkovic2004321} corresponds $\alpha=1$ and the QTDA
expression in Ref.~\cite{PhysRevC.87.064302} to $\alpha=0$.

\section{\label{sec:appen-num} Numerical Results for Matrix Elements in
Individual Nuclei}

Table \ref{table:2nbb-global1} below provides details related to Fig.\
\ref{fig:2nbb}.  Table \ref{table:2nbb-prediction} does the same for Fig.\
\ref{fig:2nbb_prediction}.

\begin{table*}[t]
\caption{Dimensionless Gamow-Teller $2\nu\beta\beta$ nuclear matrix element $m_e
c^2 M_{\rm GT}^{2\nu}$ computed with the SkO$'$- and SV-min-based EDFs, with
globally fitted proton-neutron parts \cite{PhysRevC.93.014304}.  The value of
the matrix element is compared with experimental values extracted from
Ref.~\cite{universe.6.159}.  The overlap of the intermediate states is evaluated
both with QTDA ($\alpha=0$, Ref.~\cite{PhysRevC.87.064302}) and QRPA
($\alpha=1$, Ref.~\cite{Simkovic2004321}) prescriptions.
The numbers in parentheses denote powers of 10.
\label{table:2nbb-global1}}
\begin{ruledtabular} 
  \begin{tabular}{ccccccccccccc}
    & $\alpha$ & 1A & 1B & 1C & 1D & 1E & 2 & 3A & 3B & 4 & 5 & Exp. \\ \hline
    $^{48}$Ca & 0
    &0.0759&0.0734&0.0763&0.0722&0.0934&0.0698&0.0729&0.0879&0.0706&0.118 &
    0.035$\pm$0.003\\
& 1 &0.0399&0.0386&0.0401&0.0382&0.0489&0.0372&0.0386&0.0463&0.0385&0.0588\\
$^{76}$Ge & 0
&0.0496&0.0477&0.0496&0.0441&0.062&0.0469&0.0462&0.0588&0.0502&0.0426 &
0.106$\pm$ 0.004 \\
& 1 &0.0343&0.033&0.0344&0.0304&0.0431&0.0331&0.0319&0.0409&0.0355&0.0293\\
$^{82}$Se & 0
&0.0572&0.0547&0.0572&0.05&0.0736&0.0543&0.0528&0.0693&0.0567&0.061 & 0.085$\pm$
0.001\\
& 1 &0.0464&0.0444&0.0464&0.0404&0.0599&0.0463&0.0428&0.0564&0.0474&0.0485\\
$^{96}$Zr & 0
&0.0265&0.0257&0.0267&0.026&0.0321&0.123&0.0267&0.032&0.0296&0.0228 &
0.080$\pm$0.004\\
& 1 &0.0133&0.0129&0.0134&0.013&0.0164&0.108&0.0134&0.0164&0.015&0.0113\\
($^{96}$Zr sph.) & 0
&0.202&0.197&0.204&0.202&0.243&0.123&0.207&0.244&0.232&0.173\\
& 1 &0.18&0.175&0.181&0.178&0.219&0.108&0.183&0.22&0.206&0.152\\
$^{100}$Mo & 0
&$1.71(-5)$&$1.67(-5)$&$1.72(-5)$&$1.69(-5)$&$2.02(-5)$&0.234&$1.73(-5)$&$2.01(-5)$&$1.7(-5)$&$1.53(-5)$
& 0.151$\pm$ 0.005\\
& 1
&$2.67(-6)$&$2.58(-6)$&$2.68(-6)$ &$2.57(-6)$&$3.25(-6)$&0.2&$2.67(-6)$&$3.24(-6)$&$3.16 (-6)$&$2.32(-6)$\\
($^{100}$Mo pro.) & 0
&0.166&0.161&0.167&0.162&0.201&0.234&0.166&0.199&0.179&0.147\\
& 1 &0.142&0.138&0.143&0.137&0.175&0.2&0.142&0.173&0.16&0.125\\
$^{116}$Cd & 0 &0.11&0.107&0.111&0.108&0.131&0.0606&0.11&0.13&0.082&0.0997 &
0.108$\pm$0.003\\
& 1 &0.0728&0.0708&0.0732&0.0707&0.0865&0.0132&0.0725&0.086&0.0541&0.0655\\
$^{128}$Te & 0
&0.0161&0.0153&0.0161&0.0137&0.0215&0.0124&0.0149&0.0207&0.00873&0.0131 &
0.043$\pm$0.003 \\
& 1
&0.00993&0.00944&0.00994&0.00848&0.0134&0.00695&0.00923&0.0129&0.00626&0.00808\\
($^{128}$Xe obl.) & 0
&0.0306&0.0291&0.0306&0.0263&0.0407&0.0263&0.0285&0.0393&0.0155&0.0251\\
& 1 &0.0195&0.0185&0.0195&0.0165&0.0264&0.0154&0.018&0.0253&0.0106&0.0159\\
$^{130}$Te & 0
&0.0227&0.0215&0.0227&0.0189&0.0308&0.0215&0.0208&0.0295&0.0149&0.0185
&0.0293$\pm$0.0009 \\
& 1 &0.0168&0.0159&0.0168&0.0141&0.0229&0.0151&0.0154&0.0219&0.0118&0.0138\\
$^{136}$Xe & 0
&0.0222&0.0208&0.0221&0.0173&0.0318&0.0238&0.0194&0.0296&0.0184&0.0232 &
0.0181$\pm$ 0.0006\\
& 1 &0.018&0.0169&0.018&0.0139&0.0261&0.0201&0.0156&0.0243&0.0136&0.0175\\
$^{150}$Nd & 0
&0.0413&0.0395&0.0414&0.0369&0.0541&0.0552&0.0399&0.0536&0.0511&0.0341 & 0.044
$\pm$ 0.005 \\
& 1 &0.0345&0.0329&0.0346&0.0308&0.0455&0.0463&0.0334&0.0451&0.0425&0.0284\\
$^{238}$U & 0
&0.0462&0.044&0.0462&0.039&0.0616&0.048&0.0434&0.0609&0.0717&0.0374 &
0.13$^{+0.09}_{-0.07}$\\
& 1 &0.0428&0.0407&0.0428&0.0359&0.0573&0.0431&0.04&0.0566&0.0597&0.0345\\
\end{tabular}
\end{ruledtabular}
\end{table*}
 
\begin{table*}[t]
\caption{Dimensionless Gamow-Teller $2\nu\beta\beta$ nuclear matrix element $m_e
c^2 M_{\rm GT}^{2\nu}$ for nuclei whose half lives have not been measured yet.
The column and row labels are the same as in Table~\ref{table:2nbb-global1}.
\label{table:2nbb-prediction}}
\begin{ruledtabular} 
  \begin{tabular}{ccccccccccccc}
    & $\alpha$ & 1A & 1B & 1C & 1D & 1E & 2 & 3A & 3B & 4 & 5 \\ \hline
    $^{46}$Ca & 0
    &0.0868&0.0849&0.0874&0.0874&0.1&0.00611&0.0867&0.097&0.0831&0.117\\
& 1 &0.0529&0.0518&0.0534&0.0532&0.0616&0.00332&0.0529&0.0595&0.0534&0.07\\
$^{70}$Zn & 0
&0.0848&0.0823&0.0854&0.0823&0.102&0.189&0.0831&0.0992&0.0982&0.0807\\
& 1 &0.0263&0.0254&0.0265&0.0248&0.0326&0.138&0.0252&0.0312&0.0306&0.0246\\
$^{80}$Se & 0
&0.0673&0.0646&0.0674&0.0599&0.0849&0.0343&0.0626&0.0802&0.0657&0.0677\\
& 1 &0.0505&0.0485&0.0505&0.0446&0.0638&0.0228&0.0468&0.0603&0.0507&0.0495\\
$^{86}$Kr & 0
&0.0308&0.0293&0.0308&0.0256&0.0411&0.0267&0.0274&0.0379&0.0276&0.0433\\
& 1 &0.0228&0.0217&0.0227&0.0187&0.0303&0.0208&0.0201&0.028&0.0206&0.0294\\
$^{94}$Zr & 0 &0.166&0.162&0.168&0.169&0.195&0.0959&0.171&0.196&0.195&0.146\\
& 1 &0.147&0.143&0.148&0.148&0.175&0.0841&0.151&0.175&0.172&0.127\\
$^{98}$Mo & 0 &0.186&0.181&0.188&0.186&0.223&0.171&0.189&0.222&0.208&0.169\\
& 1 &0.149&0.144&0.15&0.146&0.181&0.148&0.149&0.18&0.168&0.133\\
$^{104}$Ru & 0 &0.18&0.175&0.181&0.177&0.217&0.198&0.181&0.216&$-$0.718&0.162\\
& 1 &0.155&0.15&0.156&0.15&0.189&0.16&0.155&0.188&$-$0.149&0.137\\
$^{110}$Pd & 0 &0.179&0.173&0.18&0.175&0.215&0.187&0.18&0.214&0.279&0.16\\
& 1 &0.127&0.123&0.128&0.123&0.154&0.111&0.127&0.153&0.134&0.113\\
$^{114}$Cd & 0 &0.126&0.123&0.127&0.126&0.149&0.0506&0.128&0.148&0.102&0.115\\
& 1 &0.0771&0.0751&0.0776&0.0762&0.0911&0.00702&0.0775&0.0908&0.0623&0.0699\\
$^{122}$Sn & 0
&0.0279&0.0265&0.028&0.0239&0.0377&0.0193&0.026&0.0364&0.0645&0.0226\\
& 1 &0.0171&0.0162&0.0172&0.0145&0.0234&0.00955&0.0159&0.0225&0.0285&0.0137\\
$^{124}$Sn & 0
&0.0488&0.0462&0.0489&0.041&0.0676&0.0263&0.0451&0.065&0.055&0.0391\\
& 1 &0.0382&0.0361&0.0382&0.0319&0.053&0.0157&0.0351&0.0508&0.0386&0.0303\\
$^{134}$Xe & 0
&0.0203&0.0192&0.0202&0.017&0.0274&0.0218&0.0185&0.0261&0.0142&0.0176\\
& 1 &0.015&0.0142&0.015&0.0126&0.0204&0.016&0.0138&0.0195&0.0114&0.0131\\
$^{142}$Ce & 0
&0.0289&0.0281&0.029&0.0277&0.0339&0.0168&0.0291&0.0343&0.0322&0.025\\
& 1 &0.0224&0.0218&0.0225&0.0215&0.0264&0.013&0.0226&0.0266&0.026&0.0194\\
$^{146}$Nd & 0
&0.0117&0.0113&0.0117&0.0109&0.0145&0.0135&0.0116&0.0145&0.0145&0.00979\\
& 1
&0.00512&0.00491&0.00514&0.0047&0.00655&0.00832&0.00505&0.00654&0.00663&0.00419\\
$^{148}$Nd & 0 &0.116&0.112&0.117&0.109&0.149&0.109&0.116&0.149&0.137&0.0979\\
& 1 &0.0937&0.0898&0.0942&0.0869&0.121&0.0826&0.0928&0.121&0.112&0.0782\\
$^{154}$Sm & 0
&0.0725&0.0694&0.0728&0.0652&0.0944&0.0639&0.0701&0.0933&0.0958&0.0603\\
& 1 &0.0658&0.0629&0.066&0.0589&0.0862&0.0571&0.0635&0.0852&0.083&0.0545\\
$^{160}$Gd & 0
&0.0847&0.081&0.085&0.0759&0.11&0.0807&0.0819&0.109&0.097&0.0704\\
& 1 &0.0766&0.0732&0.0768&0.0682&0.1&0.0724&0.0737&0.0994&0.0914&0.0633\\
$^{170}$Er & 0
&0.0753&0.0722&0.0756&0.0677&0.0974&0.0844&0.0729&0.0965&$-$0.0389&0.0627\\
& 1 &0.0651&0.0622&0.0652&0.0578&0.0847&0.0612&0.0626&0.0837&0.025&0.0538\\
$^{176}$Yb & 0
&0.0657&0.0627&0.0659&0.0585&0.087&0.0391&0.0635&0.0862&$-$0.0577&0.0542\\
& 1 &0.0557&0.0531&0.0558&0.0493&0.074&0.026&0.0536&0.0733&0.0181&0.0457\\
$^{186}$W & 0
&0.0966&0.0923&0.0969&0.0853&0.127&0.0637&0.0923&0.125&$-$0.07&0.0799\\
& 1 &0.0892&0.0851&0.0894&0.0781&0.118&0.0512&0.0848&0.116&0.00665&0.0733\\
$^{192}$Os & 0
&0.0672&0.0642&0.0673&0.0579&0.0881&0.0927&0.0631&0.0863&$-$0.0446&0.0553\\
& 1 &0.0618&0.0589&0.0618&0.0529&0.0814&0.083&0.0578&0.0797&$-$0.0295&0.0506\\
$^{198}$Pt & 0
&0.0272&0.0262&0.0272&0.0238&0.0341&0.0478&0.0255&0.0333&0.396&0.0229\\
& 1 &0.0167&0.016&0.0167&0.0144&0.0209&0.0279&0.0155&0.0204&$-$0.0265&0.014\\
$^{204}$Hg & 0
&0.0133&0.0127&0.0132&0.0107&0.0167&0.0195&0.0117&0.016&$-$0.023&0.0109\\
& 1 &0.0108&0.0104&0.0107&0.0087&0.0136&0.0161&0.00954&0.013&0.0178&0.00885\\
$^{226}$Ra & 0
&0.0739&0.0703&0.074&0.064&0.0987&0.0706&0.0708&0.0986&0.343&0.06\\
& 1 &0.068&0.0646&0.068&0.0585&0.0913&0.061&0.0649&0.091&0.172&0.0549\\
$^{232}$Th & 0
&0.0509&0.0485&0.0509&0.0434&0.0678&0.0531&0.0481&0.0672&0.171&0.0414\\
& 1 &0.0465&0.0443&0.0465&0.0394&0.0623&0.0433&0.0438&0.0617&0.0894&0.0376\\
$^{244}$Pu & 0
&0.0431&0.0409&0.043&0.0359&0.0576&0.0454&0.0401&0.0568&0.0265&0.0347\\
& 1 &0.0399&0.0379&0.0399&0.0331&0.0536&0.0404&0.037&0.0528&0.0384&0.0321\\
$^{248}$Cm & 0
&0.0415&0.0394&0.0414&0.0346&0.0552&0.0435&0.0387&0.0545&$-$0.00664&0.0335\\
& 1 &0.0389&0.0369&0.0388&0.0322&0.052&0.0391&0.0361&0.0512&0.0184&0.0313
\end{tabular}
\end{ruledtabular}
\end{table*}

%\bibliographystyle{apsrev4-2}
%\bibliography{famforbbdecay}


\end{document}